\documentclass{ieeeaccess}
\usepackage{cite}
\usepackage{amsmath,amssymb,amsfonts}
\usepackage{algorithmic}
\usepackage{graphicx}
\usepackage{textcomp}

\usepackage{color}
\usepackage{soul}
\usepackage[]{graphicx}
\usepackage{amsthm}
\usepackage{amsmath}
\usepackage[varg]{txfonts}
\usepackage{bm}
\usepackage{subfigure}
\usepackage{subfigmat}
\usepackage{algorithm}
\usepackage{algorithmic}
\usepackage{array}
\usepackage{stfloats}
\usepackage{breqn}
\usepackage{threeparttable}
\usepackage{epstopdf}
\newcommand{\argmax}{\mathop{\rm{arg~max}}\limits}

\newtheorem{theorem}{Theorem}[section]
\newtheorem{proposition}{Proposition}
\usepackage[utf8]{inputenc}
\usepackage[T1]{fontenc}
\usepackage[english]{babel}
\usepackage{siunitx}

\usepackage[pagewise]{lineno}

\def\BibTeX{{\rm B\kern-.05em{\sc i\kern-.025em b}\kern-.08em
    T\kern-.1667em\lower.7ex\hbox{E}\kern-.125emX}}
\begin{document}
\history{}
\doi{}

\title{Determinant-based Fast Greedy Sensor Selection Algorithm}
\author{
\uppercase{Yuji Saito}\authorrefmark{1},
\uppercase{Taku Nonomura}\authorrefmark{1},
\uppercase{Keigo Yamada}\authorrefmark{1},
\uppercase{Kumi Nakai}\authorrefmark{1},
\uppercase{Takayuki Nagata}\authorrefmark{1},
\uppercase{Keisuke Asai}\authorrefmark{1},
\uppercase{Yasuo Sasaki}\authorrefmark{2}, and 
\uppercase{Daisuke Tsubakino}\authorrefmark{2}\IEEEmembership{Member, IEEE}
}
\address[1]{Department of Aerospace Engineering, Tohoku University, Sendai, Miyagi 980-8579, Japan}
\address[2]{Department of Aerospace Engineering, Nagoya University, Nagoya, Aichi 464-8603, Japan}
\tfootnote{This work was supported by JST ACT-X Grant Number JPMJAX20AD and JST CREST Grant Number JPMJCR1763, Japan. The second author T.N. is grateful for the support of grant JPMJPR1678 of JST Presto.}
\markboth
{Y. Saito \headeretal: Determinant-based Fast Greedy Sensor Selection Algorithm}
{Y. Saito \headeretal: Determinant-based Fast Greedy Sensor Selection Algorithm}
\corresp{Corresponding author: Yuji Saito (e-mail: yuji.saito@tohoku.ac.jp).}
\begin{abstract}
In this paper, the sparse sensor placement problem for least-squares estimation is considered, and the previous novel approach of the sparse sensor selection algorithm is extended. The maximization of the determinant of the matrix which appears in pseudo-inverse matrix operations is employed as an objective function of the problem in the present extended approach. The procedure for the maximization of the determinant of the corresponding matrix is proved to be mathematically the same as that of the previously proposed QR method when the number of sensors is less than that of state variables (undersampling). 
On the other hand, the authors have developed a new algorithm for when the number of sensors is greater than that of state variables (oversampling). Then, a unified formulation of the two algorithms is derived, and the lower bound of the objective function given by this algorithm is shown using the monotone submodularity of the objective function.
The effectiveness of the proposed algorithm on the problem using real datasets is demonstrated by comparing with the results of other algorithms. The numerical results show that the proposed algorithm improves the estimation error by approximately 10\% compared with the conventional methods in the oversampling case, where the estimation error is defined as the ratio of the difference between the reconstructed data and the full observation data to the full observation. For the NOAA-SST sensor problem, which has more than ten thousand sensor candidate points, the proposed algorithm selects the sensor positions in few seconds, which required several hours with the other algorithms in the oversampling case on a 3.40 GHz computer.
\end{abstract}


\titlepgskip=-15pt

\maketitle

\section{Introduction}
\label{sec:introduction}
\begin{figure}[t]
    \centering
    \includegraphics[width=3in]{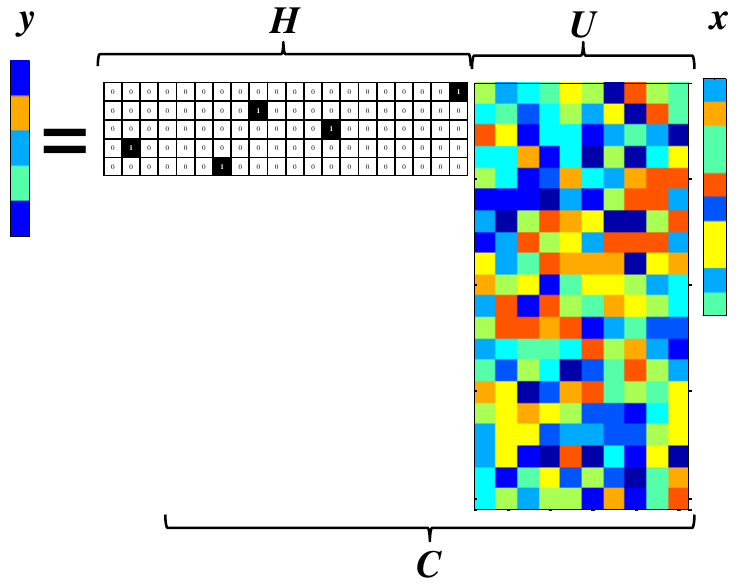}
    \caption{Graphical representation for sensor matrix $\bm{H}$ of \eqref{eq:y_cx}.}
    \label{fig:Graphicalimag}
\end{figure}

\PARstart{R}{educed-order} modeling for fluid analysis has been gathering a lot of attention. With regards to reduced-order modeling, a proper orthogonal decomposition (POD)\cite{berkooz1993proper,taira2017modal} is one of the effective methods for decomposing high-dimensional data into several significant modes of flow fields. Here, POD means a data-driven method which gives the most significant and relevant structure in the data and exactly corresponds to principal component analysis and Karhunen-Lo\`{e}ve (KL) decomposition, where the decomposed modes are orthogonal to each other. The POD analysis for a discrete data matrix can be carried out by applying singular value decomposition, as is often the case in the engineering fields. 
There are several advanced data-driven methods: dynamic mode decomposition\cite{schmid2010dynamic,Kutz2016}; empirical mode decomposition; and others that include efforts by the authors\cite{nonomura2018dynamic,nonomura2019extended}.
This research is only based on the POD which is the most basic data-driven method for reduced-order modeling.
If the data, such as flow fields, can be effectively expressed by a limited number of POD modes, limited sensors placed at appropriate positions will give the approximated full state information. Such effective observation might be one of the keys for flow control and flow prediction. This idea has been adopted by Manohar \textit{et al.}\cite{MANOHAR2018DATA}, and the sparse-sensor-placement algorithm has been developed and discussed. The idea is expressed by the following equation:
\begin{align}
\bm{y}=\bm{HUx}=\bm{C}\bm{x}.\label{eq:y_cx}
\end{align}
Here, $\bm{y}\in \mathbb{R}^p$, $\bm{x} \in \mathbb{R}^r$, $\bm{H} \in \mathbb{R}^{p\times n}$ and $\bm{U} \in \mathbb{R}^{n\times r}$ are the observation vector, the POD mode amplitude, the sparse sensor location matrix, and the spatial POD modes, respectively. In addition, $p$, $n$, and $r$ are the number of sensor locations, the degrees of freedom of the spatial POD modes, and the number of POD modes, respectively. The problem associated with the above equation is considered to be a sensor selection problem when $\bm{U}$ and strength $\bm{x}$ are assumed to be a sensor-candidate matrix and the latent state variables, respectively. A graphical representation of the equation is shown in Fig. \ref{fig:Graphicalimag}, where the element corresponding to the sensor location is unity and the others are 0 in each row of $\bm{H}$.
The sparse-sensor-placement problem is determining the optimal sensor placement of the limited $p$ sensors from $n$ sensor candidate points in this paper. Optimal sensor placement is an important challenge in the design, prediction, estimation, and control of high-dimensional systems. High-dimensional states can often leverage a latent low-dimensional representation, and this inherent compressibility enables sparse sensing. For example, in the applications of aerospace engineering, such as launch vehicles and satellites, optimal sensor placement is an important subject in performance prediction, control of the system, fault diagnostics and prognostics, etc., because of the limitations of installation, cost, and downlink capacity for transferring measurement data. Therefore, the study of optimal sensor placement is important in this field. Such sparse point sensors should be selected considering the POD modes. Although compressed sensing can recover a wider class of signals, the benefits of exploiting known patterns in data with optimized sensing is utilized, similar to as in a previous study\cite{MANOHAR2018DATA}. Drastic reductions in the required number of sensors and improved reconstruction can be expected in this case.
In a series of studies, Clark \textit{et al.} extended this idea and developed an optimization method for placing sensors under a cost constraint\cite{clark2018greedy}. Manohar \textit{et al.} developed a sensor optimization method using balanced truncation for linear systems\cite{manohar2018optimal}. Saito \textit{et al.} extended the greedy method to vector sensor problems in the context of a fluid dynamic measurement application\cite{saito2020data}.
Thus far, this sensor selection problem has been solved by convex approximation and a greedy algorithm, where the greedy algorithm was shown to be much faster than the convex approximation algorithm\cite{MANOHAR2018DATA}.

A previous study\cite{MANOHAR2018DATA} introduced a greedy algorithm based on the QR-discrete-empirical-interpolation method (QDEIM) \cite{chaturantabut2010nonlinear,drmac2016new}.
Here, QR stands for QR factorization with column pivoting matrix decomposition.
They extended the greedy algorithm for the least-squares problem when the number of sensors is greater than that of state variables.
Both convex approximation and greedy methods work reasonably well for the sensor selection problem.
Joshi and Boyd\cite{joshi2009sensor} defined the objective function for the sensor selection problem, which corresponds to minimization of the volume of the confidence ellipsoid. Joshi and Boyd\cite{joshi2009sensor} proposed a convex approximation method for this objective function. On the other hand, the proposed convex approximation method suffers from a long computational time. Recently, Nonomura \textit{et al.} extended this convex approximation method to a randomized subspace Newton convex approximation method for sparse sensor selection\cite{nonomura2021Randomized}, while Nagata \textit{et al.} proposed convex/nonconvex methods of the sensor selection problems using the alternating direction method of multipliers (ADMM)\cite{nagata2021data}. 
Although those proposed new convex methods reduce the complexity from that of the conventional convex approximation method, e.g. (${\mathcal O}(mnr^2)$) in the ADMM method compared with (${\mathcal O}(mn^3)$) in the conventional convex approximation method, they are still slower than the greedy method (${\mathcal O}(pnr^2)$) because those convex methods require $m$ iterative calculations that are usually greater than the number of sensors ($p$).
Therefore, this paper focuses on the greedy algorithm because the convex approximation method is not computationally efficient for high-dimensional data.
Manohar \textit{et al.}\cite{MANOHAR2018DATA} have proposed a QR method that provides an approximate greedy solution for the optimization, which is known to be a submatrix volume maximization, and improved the computational time of the problem, relative to the convex approximation method, by using the QR method. 
However, the QR method does not seem to connect straightforwardly with the problem $\bm{y}=\bm{Cx}$ and lacks a mathematical proof for its implementation in\cite{MANOHAR2018DATA}.
Particularly when the number of sensors is greater than that of state variables, the theoretical background for calculation is unclear for the greedy algorithm, and the complexity of the QR method is not much different from that of the convex approximation method \cite{MANOHAR2018DATA} (See Appendix \ref{Appendix:Summary of previous greedy algorithm and QR decomposition in the oversampling case} for a detailed explanation).
Therefore, the excellent idea of using the QR method is extended in this study.
Peherstorfer \textit{et al.} proposed an oversampling-point selection method in the framework of a DEIM-based reduced order model; the method is based on lower bounds of the smallest eigenvalues of certain structured matrix updates in the case of the number of sensors being greater than that of state variables\cite{peherstorfer2020stability}. They showed that their proposal is almost the best choice among the existing oversampling-point selection methods\cite{MANOHAR2018DATA,astrid2008missing,barrault2004empirical,chaturantabut2010nonlinear,carlberg2013gnat} reported in DEIM studies.

\begin{figure*}[htbp]
    \centering
    \includegraphics[width=\textwidth]{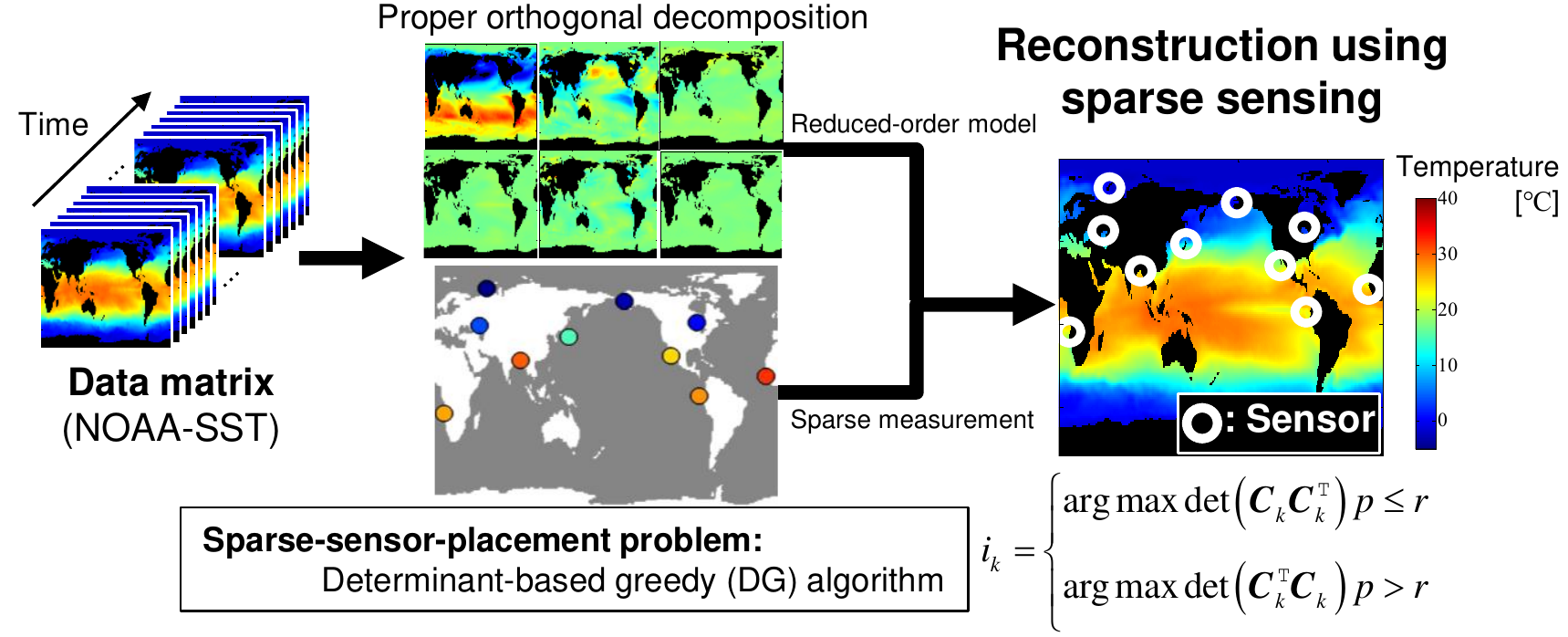}
    \caption{Conceptual flow chart of this paper (NOAA-SST sensor problem).}
    \label{fig:concept}
\end{figure*}

This paper introduces an improved formulation and proves its monotone submodularity and presents its performance. 
Fig. \ref{fig:concept} summarizes the concept of this paper the NOAA-SST sensor problem.
The main novelties and contributions of this paper are as follows:
\begin{itemize}
 \item The main contribution of this paper is extending the previous DEIM studies and the QR method to the cases of undersampled and oversampled sensor placement, where the number of sensors is less and greater than the number of latent state variables, respectively.
 \item The procedure for the maximization of the determinant of the corresponding matrix is mathematically proved to be the same as that of the QR method when the number of sensors is less than that of state variables (see Theorem \ref{theorem:QR=DG_in_the_k_less_than_p}).
 \item A unified formulation is derived, and the lower bound of the objective function is shown using monotone submodularity.
 \item A new hybrid algorithm is proposed in this paper, based on the abovementioned results. The effectiveness of the proposed new hybrid algorithm is demonstrated by comparing with the results of other algorithms, through the problems using the randomly generated data and the real datasets of the flowfield around an airfoil and the global climate. 
 \item The demonstrations show that the proposed algorithm is faster than the other methods and obtains one of the highest determinants and almost the smallest level of estimation errors.
\end{itemize}
The results for the methods proposed by Joshi and Boyd\cite{joshi2009sensor} and Manohar \textit{et al.}\cite{MANOHAR2018DATA} will be compared to our results. 
The results for the method proposed by Peherstorfer \textit{et al.} \cite{peherstorfer2020stability} are not compared in the main discussion of this paper because the objective function of \cite{peherstorfer2020stability} is different from ours. 
Instead, the superiority of the present algorithm to the method proposed by Peherstorfer \textit{et al.} \cite{peherstorfer2020stability} is demonstrated in Appendix \ref{Appendix:Comparison to previous studies in oversampling case} by applying the algorithms to a practical example.


\section{formulations and algorithms}
\subsection{Previous Greedy Algorithm~\cite{MANOHAR2018DATA}}
In the greedy algorithm based on QR decomposition for the scalar measurement problem, the ${i}$th sensor is chosen as follows:
\begin{eqnarray}
i_{k}= \argmax_{i\, \in\, \mathcal{S}\, \backslash\, \mathcal{S}_{k}} \|\bm{v}_i\|^2_2,
\end{eqnarray}
where ${i}_k$ is the index of the $k$th selected sensor. Here, we initialize $\bm{v}_i=[\begin{array}{cccc}V_{i,1} &V_{i,2} &\dots &V_{i,r}\end{array}]$ and $\{{V}_{ij}\}=\bm{V}=\bm{U}$ and $\{V_{ij}\}=\bm{V}=\bm{UU}^T$ for the $p=r$ and $p>r$ sensor conditions, respectively, as done in\cite{MANOHAR2018DATA}. Additionally, the set of indices for the sensor candidates and the subset of indices for the determined sensors are expressed as $\mathcal{S}$ $\left(=\{1,\hdots,n\}\right)$ and $\mathcal{S}_{k}$ $\left(=\{i_{1},\hdots,i_{k}\} ,\, \mathcal{S}_{k}\subset\mathcal{S}\right)$, respectively. 
Given index $i$, matrix $\bm{V}$ is pivoted and QR decomposed using the newly selected $\bm{v}_i$. Then, the next sensor is chosen for the remaining matrix. The algorithms for $p=r$ and $p>r$ sensors are QDEIM and optimized sparse sensor placement as an extension of QDEIM, respectively.
The scalar measurement problem is treated in this study, and the extension to the vector measurement problem is left for the future study. It should be noted that Saito \textit{et al.} extended the greedy algorithm based on QR decomposition to the vector measurement problem \cite{saito2020data}.

The optimization considered is maximization of the determinant of the $\bm{C}$ matrix to stably solve for the vector $\bm{x}$ in the case of $p=r$. The selection of a sensor position is based on maximizing the norm of the corresponding row vector of the sensor-candidate matrix. On the other hand, a calculation algorithm for the condition $p>r$ has already been proposed in\cite{MANOHAR2018DATA} (which corresponds to the use of $\bm{V}=\bm{UU}^{\mathrm{T}}$ instead of $\bm{V}=\bm{U}$), but the validity of the algorithm was not well clarified in\cite{MANOHAR2018DATA}. Appendix \ref{Appendix:Summary of previous greedy algorithm and QR decomposition in the oversampling case} concisely explains the work of Manohar \textit{et al.}\cite{MANOHAR2018DATA}.

\subsection{Proposed Determinant-based Greedy Algorithm}

Let $\bm{C}_{k}$ denote the $k$th sensor-candidate matrix as follows:
\begin{align}
\bm{C}_{k}=\left[\begin{array}{ccccc}\bm{u}_{i_{1}}^{\mathrm{T}} &\bm{u}_{i_{2}}^{\mathrm{T}} &\dots &\bm{u}_{i_{k-1}}^{\mathrm{T}} &\bm{u}_{i_{k}}^{\mathrm{T}}\end{array}\right]^{\mathrm{T}},
\end{align}
where ${i}_{k}$ and $\bm{u}_{i_{k}}$ are the index of the $k$th selected sensor and the corresponding row vector of the sensor-candidate matrix, respectively.
The following algorithm can be derived straightforwardly by considering the pseudo-inverse matrix of $\bm{C}$. The greedy sensor selection problem is defined such that the locations of the first-to-$(k-1)$th sensors already selected are known. Equation \eqref{eq:y_cx} can be solved as $\bm{\hat{x}}=\bm{C}^{\dagger}\bm{y}$ and divided into two cases:
\begin{align}
\bm{\hat{x}}=\left\{\begin{array}{cc}\bm{C}^{\mathrm{T}}\left(\bm{C}\bm{C}^{\mathrm{T}}\right)^{-1}\bm{y}, & p\le r, \\
\left(\bm{C}^{\mathrm{T}}\bm{C}\right)^{-1}\bm{C}^{\mathrm{T}}\bm{y}, &p>r.\end{array} \right.
\end{align}
Here, $\bm{C}$ is assumed to be a full-column-rank or full-row-rank matrix.
Our sensor selection automatically produces a full-rank observation matrix because of the procedure of the maximization of the determinant. This assumption is reasonable because the performance of a full-rank observation matrix is clearly better than that of a rank-deficient matrix. Hence, the discussion for a full-column-rank or full-row-rank matrix only is sufficient for the algorithm developed in this study.
Here, $\bm{U}$ is considered to be a full-column-rank matrix, and the existence of any unobservable subspace is not considered for simplicity.

The optimization could also be divided into the two cases of maximizing the determinants of $\bm{CC}^{\mathrm{T}}$ and $\bm{C}^{\mathrm{T}}\bm{C}$ for $p \le r$ and $p > r$, respectively. 
The maximization of the determinant of $\bm{CC}^{\mathrm{T}}$ in the case of $p \le r$ is justified by being connected to the maximization of the determinant of $\bm{C}^{\mathrm{T}}\bm{C}$, as indicated in (\ref{eq:connection}) later in this paper.
Here, the maximization of the determinant of $\bm{C}^{\mathrm{T}}\bm{C}$ in the case of $p>r$ is justified as follows. The uniform independent Gaussian noises $\mathcal{N}(\bm{0},\sigma^2\bm{I})$ are assumed to be imposed on the observation, and the estimated parameters $\bm{\hat{x}}$ can be obtained as
\begin{align}
     \bm{\hat{x}}=\left(\bm{C}^{\mathrm{T}}\bm{C}\right)^{-1}\bm{C}^{\mathrm{T}}\left(\bm{y}+\bm{Hv}\right)
     =\bm{x}+\left(\bm{C}^{\mathrm{T}}\bm{C}\right)^{-1}\bm{C}^{\mathrm{T}}\bm{Hv}. 
    \label{eq:LS_estimation in oversampling}
\end{align}
The covariance matrix of the estimation error is expressed as follows:
\begin{align}
    &{E}\left[\left(\bm{x}-\bm{\hat{x}}\right)\left(\bm{x}-\bm{\hat{x}}\right)^{\mathrm{T}}\right] \notag\\
    &={E}\left[\left(\bm{C}^{\mathrm{T}}\bm{C}\right)^{-1}\bm{C}^{\mathrm{T}}\bm{Hv}\bm{v}^{\mathrm{T}}\bm{H}^{\mathrm{T}}\bm{C}\left(\bm{C}^{\mathrm{T}}\bm{C}\right)^{-1}\right]\notag\\
    &=\sigma^2{E}\left[\left(\bm{C}^{\mathrm{T}}\bm{C}\right)^{-1}\bm{C}^{\mathrm{T}}\bm{C}\left(\bm{C}^{\mathrm{T}}\bm{C}\right)^{-1}\right] \notag\\
    &=\sigma^2\left(\bm{C}^{\mathrm{T}}\bm{C}\right)^{-1}, \label{eq:error_covarz}
\end{align}
where ${E}\left[\bm{vv}^{\mathrm{T}}\right]=\sigma^2\bm{I}$ and $\bm{HH}^{\mathrm{T}}=\bm{I}$.
Therefore, the maximizations of the determinants of $\bm{CC}^{\mathrm{T}}$ and $\bm{C}^{\mathrm{T}}\bm{C}$ are equivalent to minimizing the determinant of the error covariance matrix, resulting in minimizing the volume of the confidence ellipsoid of the regression estimates of the linear model parameters.

The first-to-$(k-1)$th sensors are already selected and it becomes the selecting problem of only the $k$th sensor in the greedy algorithm. 
The problem in the case of $k \le r$ has already been addressed by Manohar \textit{et al.}\cite{MANOHAR2018DATA}. Their algorithm selects the sensor position based on maximizing the norm of the corresponding row vector of the sensor-candidate matrix $\bm{U}$ applying the QR decomposition, as explained in\cite{MANOHAR2018DATA}, where they note that the QR decomposition provides an approximate greedy solution for the maximization of the determinant of $\bm{C}$. The equivalence of the QR pivoting decision to the greedy heuristic maximizing the determinant was previously stated in\cite{clark2018greedy} without proof.
Therefore, the mathematical background for the maximization of the determinant of $\bm{CC}^{\mathrm{T}}$ will be described in the next section. The problem in the case of $k > r$ has also been addressed in previous studies\cite{boyd2004convex,joshi2009sensor}, and the proposed objective in this study (the maximization of the determinant of $\bm{C}^{\mathrm{T}}\bm{C}$) corresponds to designing an experiment to minimize the volume of the resulting confidence ellipsoid\cite{boyd2004convex}.

The step-by-step maximization of the determinant of $\bm{C}_{k}\bm{C}_{k}^{\mathrm{T}}$ is considered using the greedy method in the case of $k \leq r$. Here, the objective is to maximize the determinant of the matrix appearing in pseudo-inverse matrix operations, leading to a minimization of confidence intervals. The matrix can be expanded as follows (see, e.g., \cite{horn2012matrix} for a detailed derivation):
\begin{align}
\det(\bm{C}_k\bm{C}_k^{\mathrm{T}}) =&
\det\left(
    \left[
        \begin{array}{c}
            \bm{C}_{k-1} \\
            \bm{u}_i
        \end{array}
    \right]
    \left[
        \begin{array}{cc}
            \bm{C}_{k-1}^{\mathrm{T}} & \bm{u}_i^{\mathrm{T}}
        \end{array}
    \right]
\right) \notag \\
=&
\det \left(
    \left[
        \begin{array}{cc}
            \bm{C}_{k-1} \bm{C}_{k-1}^{\mathrm{T}} & \bm{C}_{k-1} \bm{u}_i^{\mathrm{T}} \\
            \bm{u}_i \bm{C}_{k-1}^{\mathrm{T}} & \bm{u}_i \bm{u}_i^{\mathrm{T}}
        \end{array}
    \right]
\right) \notag \\
=&
\bm{u}_i\left(\bm{I}-\bm{C}_{k-1}^{\mathrm{T}}\left(\bm{C}_{k-1}\bm{C}_{k-1}^{\mathrm{T}}\right)^{-1}\bm{C}_{k-1}\right)\bm{u}_i^{\mathrm{T}} \notag \\
& \times \det \left( \bm{C}_{k-1} \bm{C}_{k-1}^{\mathrm{T}} \right),
\end{align}
and, therefore,
\begin{align}
i_{k}&=\argmax_{i\, \in\, \mathcal{S}\, \backslash\, \mathcal{S}_{k}}\det\left(\bm{C}_k\bm{C}_k^{\mathrm{T}}\right)\nonumber\\
&=\argmax_{i\, \in\, \mathcal{S}\, \backslash\, \mathcal{S}_{k}} \bm{u}_i\left(\bm{I}-\bm{C}_{k-1}^{\mathrm{T}}\left(\bm{C}_{k-1}\bm{C}_{k-1}^{\mathrm{T}}\right)^{-1}\bm{C}_{k-1}\right)\bm{u}_i^{\mathrm{T}}\label{eq:objklessrargmin}.
\end{align}

In the $k$th step of sensor selection of the QR (or Gram-Schmidt) method and the present method, the following equality is obtained:
\begin{align}
\bm{v}_i\bm{v}_i^{\mathrm{T}}=\bm{u}_i\left(\bm{I}-\bm{C}_{k-1}^{\mathrm{T}}\left(\bm{C}_{k-1}\bm{C}_{k-1}^{\mathrm{T}}\right)^{-1}\bm{C}_{k-1}\right)\bm{u}_i^{\mathrm{T}},
\end{align}
and, therefore, the two algorithms are equivalent.
\begin{theorem} The QR method is mathematically equivalent to the DG method in the case of $k \le r$. \label{theorem:QR=DG_in_the_k_less_than_p}
\end{theorem} This is because $\bm{I}-\bm{C}_{k-1}^{\mathrm{T}}\left(\bm{C}_{k-1}\bm{C}_{k-1}^{\mathrm{T}}\right)^{-1}\bm{C}_{k-1}=\bm{P}_{\bm{C}_{k-1}}^{\perp}$ is the projection matrix to the orthogonal complement of the already selected sensor vector space, and therefore the absolute value of the sensor-candidate matrix in the orthogonal complement space $\bm{u}_i\left(\bm{I}-\bm{C}_{k-1}^{\mathrm{T}}\left(\bm{C}_{k-1}\bm{C}_{k-1}^{\mathrm{T}}\right)^{-1}\bm{C}_{k-1}\right)\bm{u}_i^{\mathrm{T}}$ exactly corresponds to the absolute value of $\bm{v}_i$, which is the sensor-candidate vector with the sensor components already selected in the $k$th step subtracted. See Appendix \ref{Appendix:Equivalence of operations in the QR and present methods} for a detailed proof.

The following equation illustrates a somewhat efficient way to recursively calculate $\left(\bm{C}_k\bm{C}_k^{\mathrm{T}}\right)^{-1}$ when the new sensor is found and the step number $k$ is incremented:
\begin{align}
     \left( \bm{C}_{k}\bm{C}_{k}^{\mathrm{T}}\right)^{-1} =
     \left(\left[ \begin{array}{cc}\bm{C}_{k-1}\\\bm{u}_{i_k}\end{array}\right]\left[ \begin{array}{cc} \bm{C}_{k-1}^{\mathrm{T}}& \bm{u}_{i_k}^{\mathrm{T}} \end{array} \right] \right)^{-1}
     =
     \left(\begin{array}{cc}
     \bm{A} & \bm{b}^{\mathrm{T}}\\
     \bm{b} & d
 \end{array}\right),
      \label{eq:invkltr}
 \end{align}
 where
 \begin{align}
 \bm{A}&=\left(\bm{C}_{k-1}\bm{C}_{k-1}^{\mathrm{T}}\right)^{-1}\nonumber\\
 &\quad\left\{\bm{I}+\frac{\bm{C}_{k-1}\bm{u}_{i_k}^{\mathrm{T}}\bm{u}_{i_k}\bm{C}_{k-1}^{\mathrm{T}}\left(\bm{C}_{k-1}\bm{C}_{k-1}^{\mathrm{T}}\right)^{-1}}{\bm{u}_{i_k}\left(\bm{I}-\bm{C}_{k-1}^{\mathrm{T}}\left(\bm{C}_{k-1}\bm{C}_{k-1}^{\mathrm{T}}\right)^{-1}\bm{C}_{k-1}\right)\bm{u}_{i_k}^{\mathrm{T}}}\right\},\nonumber\\
 \bm{b}&=\frac{-\bm{u}_{i_k}\bm{C}_{k-1}^{\mathrm{T}}\left(\bm{C}_{k-1}\bm{C}_{k-1}^{\mathrm{T}}\right)^{-1}}{\bm{u}_{i_k}\left(\bm{I}-\bm{C}_{k-1}^{\mathrm{T}}\left(\bm{C}_{k-1}\bm{C}_{k-1}^{\mathrm{T}}\right)^{-1}\bm{C}_{k-1}\right)\bm{u}_{i_k}^{\mathrm{T}}},\nonumber\\
 d&=\frac{1}{\bm{u}_{i_k}\left(\bm{I}-\bm{C}_{k-1}^{\mathrm{T}}\left(\bm{C}_{k-1}\bm{C}_{k-1}^{\mathrm{T}}\right)^{-1}\bm{C}_{k-1}\right)\bm{u}_{i_k}^{\mathrm{T}}}.\nonumber
 \end{align}
However, the overall computational speed of the previous QR method is faster than that of the greedy algorithm using this recursive equation. The previous QR method is therefore chosen for the case of $k<r$ except for the validation in this paper. Although, the algorithm cannot be used for practical applications in a straightforward manner, it is presented herein as Alg. \ref{alg:dgr_k_le_r}.

\begin{algorithm}
\caption{Determinant-based greedy algorithm for sparse sensor placement for $k \le r$ (corresponding to the previous QR algorithm for $p=r$)}
\begin{algorithmic}
\label{alg:dgr_k_le_r}
\STATE Set sensor-candidate matrix $\bm{U}$.
\STATE $\bm{u}_{i}=\left[\begin{array}{cccc}\bm{U}_{i,1} &\bm{U}_{i,2} &\dots &\bm{U}_{i,r}\end{array}\right]$
\STATE $i \leftarrow \argmax_{i\, \in\, \mathcal{S}} \left(1 + \bm{u}_i  \bm{u}^{\mathrm{T}}\right)$
\STATE $ H_{1,i} = 1$
\STATE $\bm{C}_1 = \bm{u}_i$
\STATE $[\begin{array}{cccc}\bm{U}_{i,1} &\bm{U}_{i,2} &\dots &\bm{U}_{i,r}\end{array}] \leftarrow [0,0,\dots,0]$

\FOR{ $k =2, \dots, r$ }
      \STATE $\bm{u}_{i}=[\begin{array}{cccc}\bm{U}_{i,1} &\bm{U}_{i,2} &\dots &\bm{U}_{i,r}\end{array}]$
      \STATE $i \leftarrow \argmax_{i\, \in\, \mathcal{S}\, \backslash\, \mathcal{S}_{k}} \bm{u}_i\left(\bm{I}-\bm{C}_{k-1}^{\mathrm{T}}\left(\bm{C}_{k-1}\bm{C}_{k-1}^{\mathrm{T}}\right)^{-1}\bm{C}_{k-1}\right)\bm{u}_i^{\mathrm{T}}$  
      \STATE Update $\left(\bm{C}_{k}\bm{C}_{k}^{\mathrm{T}}\right)^{-1}$ using Eq. \eqref{eq:invkltr}
      \STATE $H_{k,i} = 1$
      \STATE $\bm{C}_{k}=\left[\begin{array}{ccccc}\bm{u}_{i_{1}}^{\mathrm{T}} &\bm{u}_{i_{2}}^{\mathrm{T}} &\dots &\bm{u}_{i_{k-1}}^{\mathrm{T}} &\bm{u}_{i_{k}}^{\mathrm{T}}\end{array}\right]^{\mathrm{T}}$
      \STATE $\left[\begin{array}{cccc}\bm{U}_{i,1} &\bm{U}_{i,2} &\dots &\bm{U}_{i,r}\end{array}\right] \leftarrow \left[0,0,\dots,0\right]$
\ENDFOR
\end{algorithmic}
\end{algorithm}

The maximization of the determinant of $\bm{C}^{\mathrm{T}}\bm{C}$ is simply considered in the case of $k \ge r$ (see, e.g., \cite{horn2012matrix} for a detailed derivation), which is commonly known as the D-optimal criterion in design of experiments.
\begin{align}
\det\left(\bm{C}_k^{\mathrm{T}}\bm{C}_k\right)&=\det\left(
\left[ \begin{array}{cc} \bm{C}_{k-1}^{\mathrm{T}}& \bm{u}_i^{\mathrm{T}} \end{array} \right] \nonumber
\left[ \begin{array}{c}\bm{C}_{k-1}\\\bm{u}_i\end{array}\right] \right)\\
&=\left(1 + \bm{u}_i\left(\bm{C}_{k-1}^{\mathrm{T}}\bm{C}_{k-1}\right)^{-1} \bm{u}_i^{\mathrm{T}}\right) \det\left(\bm{C}_{k-1}^{\mathrm{T}}\bm{C}_{k-1} \right).
\end{align}
Therefore, 
\begin{align}
i_{k}&=\argmax_{i\, \in\, \mathcal{S}\, \backslash\, \mathcal{S}_{k}} \det\left(\bm{C}_{k-1}^{\mathrm{T}}\bm{C}_{k-1} + \bm{u}_i^{\mathrm{T}}\bm{u}_i\right)\nonumber\\
&=\argmax_{i\, \in\, \mathcal{S}\, \backslash\, \mathcal{S}_{k}} \left(1 + \bm{u}_i\left(\bm{C}_{k-1}^{\mathrm{T}}\bm{C}_{k-1}\right)^{-1} \bm{u}_i^{\mathrm{T}}\right).
\end{align}
The complexity of this procedure is only $O(r^2)$ for one component of a candidate sensor vector when $\left(\bm{C}_{k-1}^{\mathrm{T}}\bm{C}_{k-1}\right)^{-1}$ is known, and the complexity when searching all components of the vector becomes $O(nr^2)$. When a new sensor is found and the step number $k$ is incremented, $\left(\bm{C}_{k}^{\mathrm{T}}\bm{C}_{k}\right)^{-1}$ can be computed recursively using a matrix inversion lemma as follows:
\begin{align}
&\left(\bm{C}_{k}^{\mathrm{T}}\bm{C}_{k}\right)^{-1}\nonumber\\
&=\left(\bm{C}_{k-1}^{\mathrm{T}}\bm{C}_{k-1}+\bm{u}_{i}^{\mathrm{T}}\bm{u}_{i}\right)^{-1} \nonumber \\
&=\left(\bm{C}_{k-1}^{\mathrm{T}}\bm{C}_{k-1}\right)^{-1}\nonumber\\
&\quad \left(\bm{I}-\bm{u}_{i}^{\mathrm{T}}\left(1+\bm{u}_{i}\left(\bm{C}_{k-1}^{\mathrm{T}}\bm{C}_{k-1}\right)^{-1}\bm{u}_{i}^{\mathrm{T}}\right)^{-1}\bm{u}_{i}\left(\bm{C}_{k-1}^{\mathrm{T}}\bm{C}_{k-1}\right)^{-1}\right).
\end{align}
The determinant-based greedy algorithm for sparse sensor placement for $k>r$ is summarized as Alg. \ref{alg:dgr_k_gt_r}.

\begin{algorithm}
\caption{Determinant-based greedy algorithm for sparse sensor placement for $k>r$ }
\begin{algorithmic}
\label{alg:dgr_k_gt_r}
\STATE Set sensor-candidate matrix $\bm{U}$.
\STATE Obtain $\bm{H}_{\text{o}}$ using the previous QR algorithm (or Alg. \ref{alg:dgr_k_le_r}) for the number of sensors $p_{\text{o}}=r$.
\STATE Set $\left(\bm{C}_{p_{\text{o}}}^{\mathrm{T}}\bm{C}_{p_{\text{o}}}\right)^{-1}=\left(\bm{U}^{\mathrm{T}}\bm{H}_{\text{o}}^{\mathrm{T}}\bm{H}_{\text{o}}\bm{U}\right)^{-1}$
\FOR{ $i$ of the sensor position chosen by the conventional greedy algorithm }
\STATE Set $\bm{U}_{i}=\left[\begin{array}{cccc}\bm{U}_{i,1} &\bm{U}_{i,2} &\dots &\bm{U}_{i,r}\end{array}\right]=[0,0,\dots,0]$
\ENDFOR
\FOR{ $k =p_{\text{o}}+1, \dots, p$ }
      \STATE $\bm{U}_{i}=\left[\begin{array}{cccc}\bm{U}_{i,1} &\bm{U}_{i,2} &\dots &\bm{U}_{i,r}\end{array}\right]$
      \STATE $i \leftarrow \argmax_i \left(1 + \bm{u}_i\left(\bm{C}_{k-1}^{\mathrm{T}}\bm{C}_{k-1}\right)^{-1} \bm{u}_i\right)$
      \STATE $\left(\bm{C}_{k}^{\mathrm{T}}\bm{C}_{k}\right)^{-1}$ \STATE$\hspace{15pt}\leftarrow \left(\bm{C}_{k-1}^{\mathrm{T}}\bm{C}_{k-1}\right)^{-1}$\\
      $\quad\quad\left(I-\bm{u}_{i}^{\mathrm{T}}\left(1+\bm{u}_{i}\left(\bm{C}_{k-1}^{\mathrm{T}}\bm{C}_{k-1}\right)^{-1}\bm{u}_{i}^{\mathrm{T}}\right)^{-1}\bm{u}_{i}\left(\bm{C}_{k-1}^{\mathrm{T}}\bm{C}_{k-1}\right)^{-1}\right)$ 
      \STATE $H_{k,i} = 1$
       \STATE $\bm{C}_{k}=[\begin{array}{ccccc}\bm{u}_{i_{1}}^{\mathrm{T}} &\bm{u}_{i_{2}}^{\mathrm{T}} &\dots &\bm{u}_{i_{k-1}}^{\mathrm{T}} &\bm{u}_{i_{k}}^{\mathrm{T}}\end{array}]^{\mathrm{T}}$
      \STATE $[\begin{array}{cccc}\bm{U}_{i,1} &\bm{U}_{i,2} &\dots &\bm{U}_{i,r}\end{array}] \leftarrow [0,0,\dots,0]$
\ENDFOR
\end{algorithmic}
\end{algorithm}
\subsection{Unified Expressions for Proposed Algorithm} 
\label{sub-sec:Unified Expression for Proposed Algorithm}
Here, we introduce unified expressions for the case in which the number of sensors is less than that of modes and that in which the number of sensors is greater than or equal to that of modes. The maximization of the term $\det\left(\bm{C}^{\mathrm{T}}\bm{C} + \epsilon \bm{I}\right)$ is considered in this case, where $\epsilon$ is a sufficiently small number. This objective function approximately corresponds to $\det\left(\bm{C}^{\mathrm{T}}\bm{C}\right)$ when the number of sensors is greater than or equal to that of the modes. 
On the other hand, when the number of sensors $p$ is less than that of the modes $r$, it becomes
\begin{align}
\det\left(\bm{C}^{\mathrm{T}}\bm{C} + \epsilon \bm{I}\right)&=\epsilon^r\det 
\left(\bm{I} + \frac{1}{\epsilon}\bm{C}^{\mathrm{T}}\bm{C}  \right)\nonumber \\ 
&=\epsilon^r\det 
\left(\bm{I} + \frac{1}{\epsilon}\bm{C}\bm{C}^{\mathrm{T}}  \right)\nonumber \\
&=\epsilon^{r-p}\det 
\left(\bm{C}\bm{C}^{\mathrm{T}} + \epsilon \bm{I} \right),
\label{eq:connection}
\end{align}
and then the objective corresponds to the maximization of $\det 
\left(\bm{C}\bm{C}^{\mathrm{T}} + \epsilon \bm{I} \right) \sim \det 
\left(\bm{C}\bm{C}^{\mathrm{T}}  \right) $
as we did in the previous subsection. Therefore, the approximated objective function is defined to be $\det(\bm{C}^{\mathrm{T}}\bm{C} + \epsilon \bm{I})$ which asymptotically approaches the maximum of the determinant in the limit that $\epsilon$ goes to $0$. Although this function with sufficiently small $\epsilon$ should not be implemented in the computational code because it may include large round-off error when $\epsilon$ is sufficiently small to neglect its effects, this unified function will be used for the proof of monotone submodularity and discussion of the approximation rate in the next section. 

\section{Submodularity and approximation rate}\label{sec:Unified Expressions for Proposed Algorithm} 
It is well known that an explicit approximation rate of the greedy algorithm is available if the objective function is monotone and submodular \cite{nemhauser1978analysis}. 
In this section, we show that a slightly modified version of our unified objective function is certainly monotone and submodular.
By using the approximation rate for this function, we derive a lower bound for the proposed algorithm.

Let $S \subset \{ 1, 2, \dots, n \}$ be a set of labels of selected sensors and $\bm{C}_{S}$ be the corresponding sensor matrix.
Namely, if $S = \{ i_{1}, i_{2}, \dots, i_{k} \}$, then $\bm{C}_{S}$ is given by
\begin{align}
 \bm{C}_{S} 
 = \left[ 
      \begin{array}{@{\,}cccc@{\,}}
        \bm{u}_{i_{1}}^{\mathrm{T}} &
        \bm{u}_{i_{2}}^{\mathrm{T}} & \cdots &
        \bm{u}_{i_{k}}^{\mathrm{T}}
      \end{array}
   \right]^{\mathrm{T}}.
\end{align}
Define a function $f: 2^{ \{ 1, 2, \dots, n \} } \to \mathbb{R} $ by
\begin{align}
 f(S) 
&= \log \det ( \bm{C}_{S}^{\mathrm{T}} \bm{C}_{S}^{} + \epsilon \bm{I} )
   - r \log \epsilon 
 \notag  \\
&= \log 
   \det \left( 
          \frac{1}{\epsilon} 
          \bm{C}_{S}^{\mathrm{T}} \bm{C}_{S}^{} + \bm{I} 
        \right)
 \label{eq:def_f}
\end{align}
for each $S \subset \{ 1, 2, \dots, n \}$.
An offset term $-r \log \epsilon$ is added so that the value of $f$ for the empty set can be regarded as $f(\emptyset) = \log \det \bm{I} = 0$.

In what follows, we first show that $f$ is a submodular function.
Namely, for any $S, T \subset \{ 1, 2, \dots, n \} $ with $S \subset T$ and $i \in \{ 1, 2, \dots, n \} \setminus T$, the function $f$ satisfies
\begin{align}
     f\bigl( S \cup \{ i \} \bigr) - f(S)
 \ge f\bigl( T \cup \{ i \} \bigr) - f(T).
 \label{eq:def_submodular_functions}
\end{align}
The monotonicity of $f$ is next proved.

\begin{proposition}
The function 
 $f : 2^{\{ 1, 2, \dots, n \}} \to \mathbb{R}$ defined by \eqref{eq:def_f}
 is submodular. 
 \label{thm:submodularity}
\end{proposition}

\begin{proof}
Take arbitrary $S, T \subset \{ 1, 2, \dots, n \}$ such that $S \subset T$.
 For simplicity of notation, let us set 
 $\bm{\Psi}_{\epsilon} := (\bm{C}_{S}^{\mathrm{T}} \bm{C}_{S}^{} + \varepsilon \bm{I})^{-1}$.
 It follows from the definition of $f$ that, for any 
 $i \in \{ 1, 2, \dots, n\} \setminus T,$
 \begin{align}
 &  f (S \cup \{i\}) - f(S) \notag \\
 &= \log \det \bigl(
                \bm{C}_S^{\mathrm{T}}\bm{C}_S^{} + \bm{u}_i^{\mathrm{T}} \bm{u}_i^{}
                + \epsilon \bm{I}
              \bigr) 
  - \log \det \bigl( \bm{C}_S^{\mathrm{T}} \bm{C}_S^{} + \epsilon \bm{I} \bigr) \notag \\
 &= \log \det \Bigl(
                 \bigl(
                     \bm{C}_S^{\mathrm{T}} \bm{C}_S^{} + \bm{u}_i^{\mathrm{T}} \bm{u}_i^{} + \epsilon \bm{I}
                 \bigr)
                 \bm{\Psi}_{\epsilon}
              \Bigr)
     \notag \\
 &= \log \det \Bigl(
                  \bm{I} 
                  + \bm{u}_i^{\mathrm{T}} \bm{u}_i 
                    \bm{\Psi}_{\epsilon}
              \Bigr) 
    \notag \\
 &= \log 
    \Bigl(
        1 + \bm{u}_i 
            \bm{\Psi}_{\epsilon}
            \bm{u}_i^{\mathrm{T}} 
    \Bigr).
 \label{eq:submodular_Si-S}
 \end{align}
 Due to the positive definiteness of $\bm{\Psi}_{\epsilon}$, the value of \eqref{eq:submodular_Si-S} is positive for any $\bm{u}_{i} \neq 0$. 

We next evaluate $f(T \cup \{ i \}) - f(T)$. Since $S \subset T,$ there exists a permutation matrix $\bm{P}$ such that
 \begin{equation}
   \bm{P} \bm{C}_{T} 
   = \left[  
        \begin{array}{@{\,}c@{\,}}
	  \bm{C}_{S} \\ \bm{C}_{T \setminus S} 
	\end{array}
     \right].
 \end{equation}
 Hence, we have
 \begin{equation}
   \bm{C}_{T}^{\mathrm{T}} \bm{C}_{T}^{}
   = \bm{C}_{T}^{\mathrm{T}} \bm{P}^{\mathrm{T}} \bm{P} \bm{C}_{T}^{}
   = \bm{C}_{S}            ^{\mathrm{T}} \bm{C}_{S}^{} 
   + \bm{C}_{T \setminus S}^{\mathrm{T}} \bm{C}_{T \setminus S}^{},
   \label{eq:CT^TxCT}
 \end{equation}
 where the fact $\bm{P}^{\mathrm{T}} \bm{P} = \bm{I}$ has been used.
 Direct computation together with \eqref{eq:CT^TxCT} gives
 \begin{align}
 &  f (T \cup \{i\}) - f(T) \notag \\
 &= \log \det \bigl(
                \bm{C}_T^{\mathrm{T}}\bm{C}_T^{} + \bm{u}_i^{\mathrm{T}} \bm{u}_i^{}
                + \epsilon \bm{I}
              \bigr) 
  - \log \det \bigl( \bm{C}_T^{\mathrm{T}} \bm{C}_T^{} + \epsilon \bm{I} \bigr) \notag \\
 &= \log \biggl(
            1 
            + \bm{u}_i^{}
              \bigl(
                 \bm{C}_T^{\mathrm{T}}\bm{C}_T^{}
                 + \epsilon \bm{I}
              \bigr)^{-1}
              \bm{u}_i^{\mathrm{T}}
          \biggr)
  \notag \\
 &= \log \biggl(
            1 
            + \bm{u}_i^{}
              \bigl(
                  \bm{C}_{S}            ^{\mathrm{T}} \bm{C}_{S}^{} 
                + \bm{C}_{T \setminus S}^{\mathrm{T}} \bm{C}_{T \setminus S}^{}
                + \epsilon \bm{I}
              \bigr)^{-1}
              \bm{u}_i^{\mathrm{T}}
          \biggr).
 \label{eq:submodular_Ti-T_pre}
 \end{align} 
 The matrix in parentheses can be calculated as
 \begin{align}
&  \bigl(
       \bm{C}_{S}            ^{\mathrm{T}} \bm{C}_{S}^{} 
     + \bm{C}_{T \setminus S}^{\mathrm{T}} \bm{C}_{T \setminus S}^{}
     + \epsilon \bm{I}
   \bigr)^{-1} 
   \notag \\
&= \bm{\Psi}_{\epsilon}
   \Bigl(
      \bm{I}
    + \bm{C}_{T \setminus S}^{\mathrm{T}} \bm{C}_{T \setminus S}^{}
      \bm{\Psi}_{\epsilon}
   \Bigr)^{-1}   
   \notag \\ 
&= \bm{\Psi}_{\epsilon}
   \biggl(
      \bm{I}
    - \bm{C}_{T \setminus S}^{\mathrm{T}} 
      \left(
          \bm{I} 
        + \bm{C}_{T \setminus S}^{}
          \bm{\Psi}_{\epsilon}
          \bm{C}_{T \setminus S}^{\mathrm{T}}
      \right)^{-1}
      \bm{C}_{T \setminus S}^{}
      \bm{\Psi}_{\epsilon}
   \biggr)
   \notag \\
&= \bm{\Psi}_{\epsilon}
 - \bm{\Psi}_{\epsilon}
   \bm{C}_{T \setminus S}^{\mathrm{T}} 
   \left(
       \bm{I} 
     + \bm{C}_{T \setminus S}^{}
       \bm{\Psi}_{\epsilon}
       \bm{C}_{T \setminus S}^{\mathrm{T}}
   \right)^{-1}
   \bm{C}_{T \setminus S}^{}
   \bm{\Psi}_{\epsilon}
 \notag \\
&=: \bm{\Psi}_{\epsilon} - \overline{\bm{\Psi}}_{\epsilon}
 \end{align}
 Note that $\overline{\bm{\Psi}}_{\epsilon}$ is a positive semidefinite matrix.
 Substituting the above relation into \eqref{eq:submodular_Ti-T_pre} yields
 \begin{align}
  f (T \cup \{i\}) - f(T) 
  = \log \Bigl(
             1 + \bm{u}_{i}^{} \bm{\Psi}_{\epsilon} \bm{u}_{i}^{\mathrm{T}}
             - \bm{u}_{i}^{} \overline{\bm{\Psi}}_{\epsilon} \bm{u}_{i}^{\mathrm{T}}
          \Bigr).
    \label{eq:submodular_Ti-T}
 \end{align}
 Since $\overline{\bm{\Psi}}_{\epsilon}$ is positive semidefinite, the monotonicity of the logarithm gives
 \begin{align}
  \log \Bigl(
           1 + \bm{u}_{i}^{} \bm{\Psi}_{\epsilon} \bm{u}_{i}^{\mathrm{T}}
        \Bigr)
  & \ge
  \log \Bigl(
           1 + \bm{u}_{i}^{} \bm{\Psi}_{\epsilon} \bm{u}_{i}^{\mathrm{T}}
             - \bm{u}_{i}^{} \overline{\bm{\Psi}}_{\epsilon} \bm{u}_{i}^{\mathrm{T}}
        \Bigr),
 \end{align}
 which immediately implies \eqref{eq:def_submodular_functions}.
 This completes the proof.
\end{proof}

The next result guarantees the monotonicity of $f$.
\begin{proposition}
The function $f : 2^{\{ 1, 2, \dots, n \}} \to \mathbb{R}$ 
 defined by \eqref{eq:def_f} is monotone.
\end{proposition}

\begin{proof}
For given $S, T \subset \{ 1, 2, \dots, n \}$ with $S \subset T$,
 it is clear that \eqref{eq:submodular_Si-S} holds also for 
 any $i \in T \setminus S $.
 Hence, $f(S \cup \{ i \}) - f(S) \ge 0$.
 Similarly, we can show that 
 $f(S \cup \{ i \} \cup \{ j \}) - f(S) \ge 0$
 for $j \in T \setminus  (S \cup \{ i\}) $.
 Repeated application of this argument yields
 $f(S) \le f(T)$,
 which is the desired conclusion.
\end{proof}
We have seen that $f$ is a monotone submodular function.
For optimization problems with respect to monotone submodular 
functions, Nemhauser \textit{et al.} have proved that
the following inequality holds \cite{nemhauser1978analysis}:
\begin{align}
  f(S_{\mathrm{greedy}})
& \ge
  \left(
    1 - \left(
           1 - \frac{1}{k}
        \right)
  \right)^{k}
  f(S_{\mathrm{opt}}) 
  \notag \\
& \ge
  \left(
    1 - \frac{1}{e}
  \right)
  f(S_{\mathrm{opt}}),
\end{align}
where $S_{\mathrm{opt}}$ is an optimal solution and 
$S_{\mathrm{greedy}}$ is the solution obtained by the greedy algorithm.
Applying this inequality to $f$ defined by \eqref{eq:def_f} gives
\begin{align}
&\left( 
    1 - \frac{1}{e}
 \right) 
 \left(
    \log \det 
    \left( 
       \bm{C}_{S_{\mathrm{opt}}}^{\mathrm{T}}
       \bm{C}_{S_{\mathrm{opt}}}^{}
       + \epsilon \bm{I}
    \right)
    - 
    r \log \epsilon
 \right)
 \notag \\
&\le
 \log \det 
 \left( 
    \bm{C}_{S_{\mathrm{greedy}}}^{\mathrm{T}}
    \bm{C}_{S_{\mathrm{greedy}}}^{}
    + \epsilon \bm{I}
 \right)
 - r \log \epsilon.
\end{align}
Several properties of the logarithm lead to
\begin{align}
 \epsilon^{ \frac{r}{e}}
 \left(
    \det 
    \left( 
        \bm{C}_{S_{\mathrm{opt}}}^{\mathrm{T}}
        \bm{C}_{S_{\mathrm{opt}}}^{}
        + \epsilon \bm{I}
     \right) 
 \right)^{1 - \frac{1}{e} }
 \le
 \det 
 \left( 
    \bm{C}_{S_{\mathrm{greedy}}}^{\mathrm{T}}
    \bm{C}_{S_{\mathrm{greedy}}}^{}
    + \epsilon \bm{I}
 \right). 
 \label{eq:lower_bound}
\end{align}
This inequality provides the lower bound of the greedy approximation for our problem. 

Parameter $\epsilon$ was originally assumed to be sufficiently small. 
However, the coefficient $\epsilon^{r/e}$ on the right-hand side of \eqref{eq:lower_bound} tends to $0$ as $\epsilon$ goes to $0$. 
Hence, a finite $\epsilon$ should be appropriately adopted to obtain a practical lower bound for the unified greedy algorithm proposed in this paper.

\section{Results and Discussion}
The numerical experiments are conducted and the proposed methods are validated. Hereinafter, three different implementations of the greedy methods are compared: the QR, DG, and QD methods listed in Table \ref{table:Calculation method}. Here, the QR method is the greedy method proposed in\cite{MANOHAR2018DATA}, the DG method is the greedy method for the pure maximization of the determinant, and the QD method is the greedy method, one part of which is replaced by the QR method which is mathematically proved to be equivalent to the greedy method for the pure maximization of the determinant. Table \ref{table:Computational cost for selecting $p$th sensors} explains the computational complexities of each method for selecting $p$ sensors. The QD method is included because the QR implementation for the previously proposed greedy optimization is much faster than the DG method, and the two implementations are demonstrated to give us the same numerical optimized solution as each other except for round-off errors in practical situations. The random selection and convex approximation methods\cite{joshi2009sensor} are evaluated as the references, in addition to the greedy methods.

\begin{threeparttable}[ht]
\small
\centering
\caption{Greedy sensor selection methods investigated in this study} 
\begin{tabular}[t]{lll}
\hline\hline
Name & Method &Implementation\\
\hline
QR\cite{MANOHAR2018DATA}& ${p \le r}$ : QR for ${\bm{C}}$&Alg. \ref{alg:previous}\\
& ${p>r}$ : QR for ${\bm{CC}^{\mathrm{T}}}$&Alg. \ref{alg:previous}\\
\hline
DG (present)\tnote{a}& ${p \le r}$ :${\argmax\det\left(\bm{CC}^{\mathrm{T}}\right)}$&Alg. \ref{alg:dgr_k_le_r}\\
& ${p>r}$ :${\argmax\det\left(\bm{C}^{\mathrm{T}} \bm{C}\right)}$&Alg. \ref{alg:dgr_k_gt_r}\\
\hline
QD (present,& ${p \le  r}$ :QR for ${\bm{C}}$&Alg. \ref{alg:previous}\\
recommended)\tnote{a}& ${p>r}$ :${\argmax\det\left(\bm{C}^{\mathrm{T}}\bm{C}\right)}$&Alg. \ref{alg:dgr_k_gt_r}\\
\hline\hline
    \label{table:Calculation method}
\end{tabular}    
\begin{tablenotes}[para,flushleft,online,footnotesize] 
\tiny
\item[a] The DG and QD methods are mathematically equivalent algorithms as proved in this paper.
\end{tablenotes}    
\end{threeparttable}

\begin{threeparttable}[ht]
\small
\centering
\caption{Computational complexities of each method for selecting the $p$th sensor.} 
\label{table:Computational cost for selecting $p$th sensors}
\begin{tabular}[t]{ll}
\hline\hline
Name&Computational complexity\\
\hline\hline
QR\cite{MANOHAR2018DATA}& ${p \le r}$ : ${\mathcal O}\left(nr^2\right)$\\
&${p>r}$ : ${\mathcal O}\left(n^3 \right)$\\
\hline
DG~(present)& ${p \le r}$ :${\mathcal O}\left(pnr^2\right)$\\
&${p>r}$ :${\mathcal O}\left(pnr^2 \right)$\\
\hline
QD~(present,& ${p \le  r}$ :${\mathcal O}\left(pnr\right)$\\
recommended)&${p>r}$ :${\mathcal O}\left(pnr^2\right)$\\
\hline
convex\cite{joshi2009sensor}& ${\mathcal O}\left(n^{3} \right)$ per iterations\\
&*usually converges around 100 iterations.\\
\hline\hline
\end{tabular}    
\end{threeparttable}

\begin{threeparttable}[ht]
\small
\centering
\caption{Computational environments} 
\label{table:Computational enviroment}
\begin{tabular}[t]{ll}
\hline\hline
Processor information&Intel(R) Core(TM)\\
&i7-6800K@ 3.40 GHz\\
Random access memory&128 GB\\
System type&64 bit operating system\\
&x64 base processor\\
\hline
Program code &Matlab R2013a\\
& \cite{saito2019determinantbaseda}\\
\hline
Operating system&Windows 10 Pro\\
&Version:1890\\
\hline\hline
\end{tabular}    
\end{threeparttable}

\subsection{Random sensor problem}\label{subsec:Random sensor problem}
\begin{figure}[tb]
    \centering
    \includegraphics[width=3in]{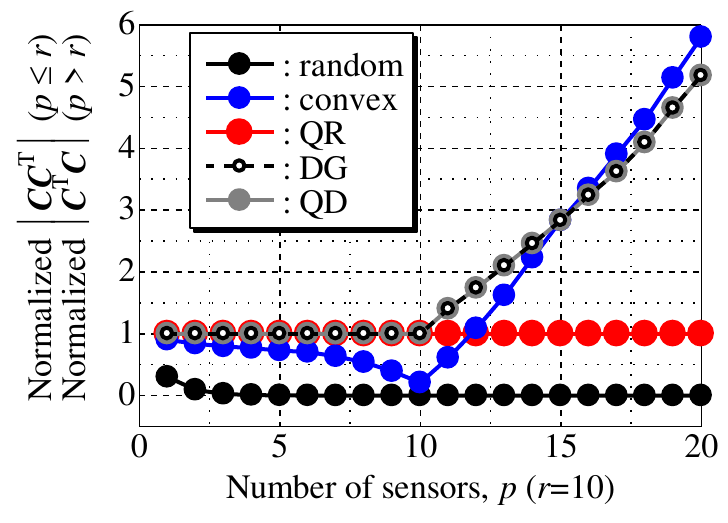}
    \caption{Normalized determinant of $\bm{CC}^{\mathrm{T}}$ ($p \le r$) or $\bm{C}^{\mathrm{T}}\bm{C}$ ($p > r$) against the number of sensors for the number of POD modes $r=10$ in the random sensor problem.}
    \label{fig:Random_normirized}
\end{figure}

The random sensor-candidate matrices, $\bm{U}\in \mathbb{R}^{2000{\times}r}$, are set according to the Gaussian distribution of $\mathcal{N}(0,1)$ with $n=2000$ and $r=10$ for the components in this validation.
The random selection, convex approximation, QR, DG, and QD methods listed in Table \ref{table:Calculation method} are evaluated. The determinants of $\bm{CC}^{\mathrm{T}}=\bm{HUU}^{\mathrm{T}}\bm{H}^{\mathrm{T}}$ ($p \le r$) and $\bm{C}^{\mathrm{T}}\bm{C}=\bm{U}^{\mathrm{T}}\bm{H}^{\mathrm{T}}\bm{HU}$ ($p > r$) are calculated using both sensor-candidate matrices after selecting sensors. The random sensor problem is solved under the computational environments listed in Table \ref{table:Computational enviroment}. The results are averages over 1000 random samples.
Fig. \ref{fig:Random_normirized} shows the relationship between the determinant of $\bm{CC}^{\mathrm{T}}$ ($p \le r$) or $\bm{C}^{\mathrm{T}}\bm{C}$ ($p > r$) and the number of sensors $p$ for the number of POD modes $r=10$ in the random sensor problem, obtained by the random selection (solid black line with closed circles), convex approximation (solid blue line with closed circles), QR (solid red line with closed circles), DG (dotted black line with open circles), and QD (solid gray line with closed circles) methods. All plots are normalized by the determinant of $\bm{CC}^{\mathrm{T}}$ ($p \le r$) or $\bm{C}^{\mathrm{T}}\bm{C}$ ($p > r$) for the QR method. This normalization shows a clear trend of each method compared with that of the conventional QR method. There exist pathological matrices for which the determinant does not increase, namely, the Kahan matrices\cite{gu1996efficient}. However, we did not experience such failures, and all the determinants of $\bm{CC}^{\mathrm{T}}$ ($p \le r$) or $\bm{C}^{\mathrm{T}}\bm{C}$ ($p > r$) increase in the random problem.
The normalized determinants of $\bm{CC}^{\mathrm{T}}$ of both the DG and QD methods are the same as that of the QR method in the case of $p \le r$, which confirms what is shown in Appendix \ref{Appendix:Equivalence of operations in the QR and present methods}. On the other hand, the normalized determinants of $\bm{C}^{\mathrm{T}}\bm{C}$ of both the DG and QD methods increase as the number of sensors increases in the case of $p > r$. 
The DG method in the case of $p>r$ selects the sensor positions using the step-by-step maximization of the determinant of $\bm{C}^{\mathrm{T}}\bm{C}$. On the other hand, the QR method conducts the full QR decomposition for $\bm{UU}^{\mathrm{T}}$ and does not seem to connect straightforwardly with the maximization of the determinant of $\bm{C}^{\mathrm{T}}\bm{C}$. For example, the determinants of $\bm{C}^{\mathrm{T}}\bm{C}$ of both the DG and QD methods are five times higher than that of the QR method in the case of $p=20$ for the random sensor problem.
Although the computational complexity or time of the convex approximation method is much larger than those of other calculation methods, the convex approximation method obtains the highest normalized determinant of $\bm{C}^{\mathrm{T}}\bm{C}$ for $p>15$ among these methods. 

\begin{figure}[htbp]
    \centering
    \includegraphics[width=3in]{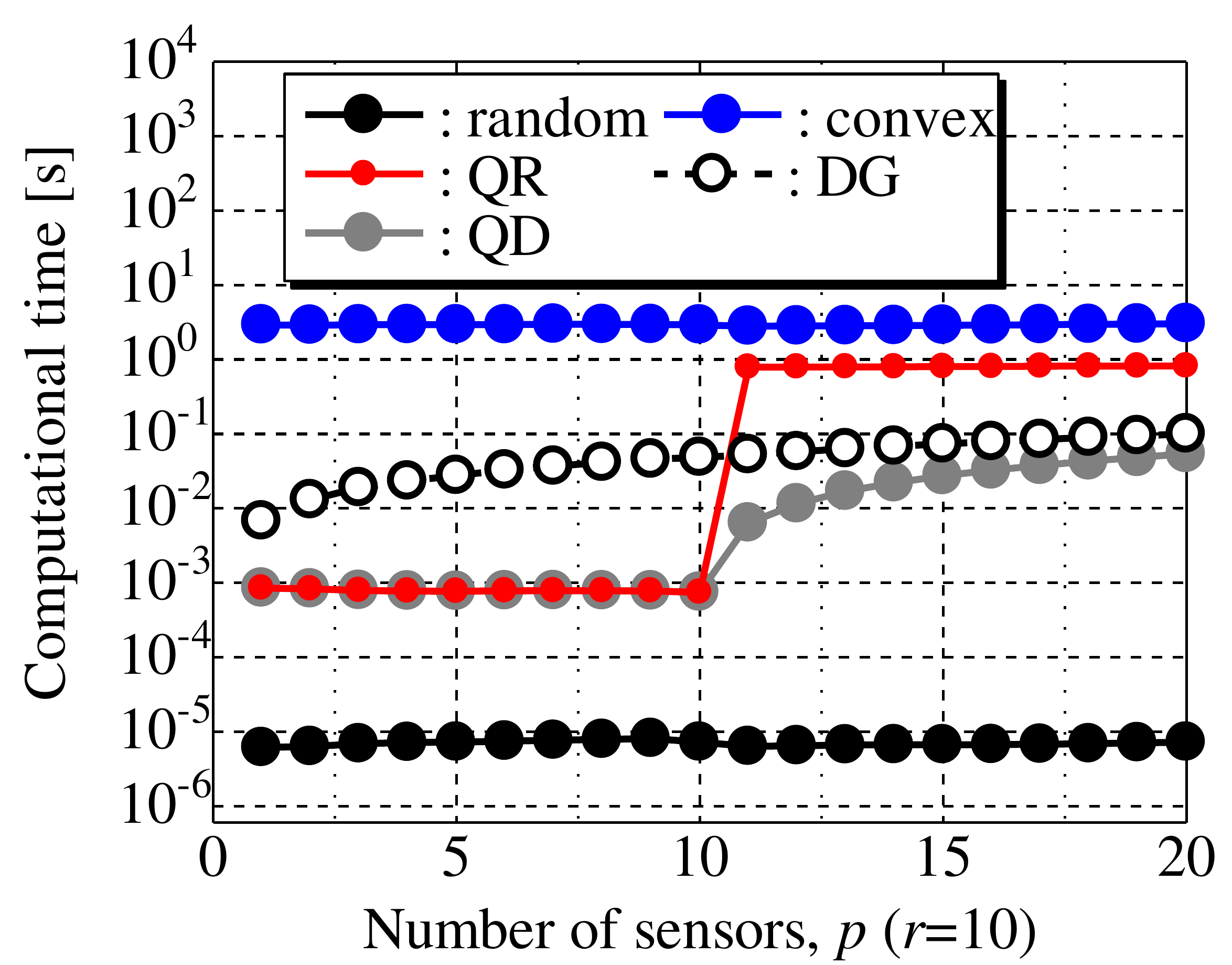}
    \caption{Computational time against the number of sensors for the number of POD modes $r=10$ in the random sensor problem.}
    \label{fig:Random_time}
\end{figure}

Fig.~\ref{fig:Random_time} shows the relationship between the computational time and the number of sensors $p$ for the number of POD modes $r=10$ in the random sensor problem. As indicated in Table~\ref{table:Computational cost for selecting $p$th sensors}, the computational complexity of the convex approximation algorithm is the largest among these calculation methods. The computational complexity of the QR method is close to that of the convex approximation method because the QR method is implemented in Matlab using the native Matlab ``qr'' subroutine and returns the desired list of all sensors.
The QR and QD methods for $p \le r$ give the same time because the same algorithm is used in this case. The computational time of the DG method gradually increases as the number of sensors increases because each sensor is chosen using the step-by-step maximization of the determinant of $\bm{C}_{k}\bm{C}_{k}^{\mathrm{T}}$ or $\bm{C}_{k}^{\mathrm{T}}\bm{C}_{k}$. On the other hand, the computational time for the QR method is almost constant for $1\le p\le 10$ and $11\le p\le 20$, because it takes a long time to conduct the full QR decomposition. The computational time of the QR method has a large difference at $p=11$ because the dimension of the matrix processed by the QR decomposition increases between the computation for $\bm{U}\in \mathbb{R}^{n{\times}r}$ and that for $\bm{UU}^{\mathrm{T}}\in \mathbb{R}^{n{\times}n}$. 
The trend of the computational time for the QD method is similar to that for the QR method, but the computational time of the QD method is much shorter than that of the QR method in the case of $p>r$. 
For example, the computational time of the QD method is approximately 10 times smaller than that of the QR and convex methods in the case of $p=20$ for the random sensor problem.
Therefore, the QD method proposed in this paper is more effective for sparse sensor placement than the convex algorithm because the determinants of $\bm{CC}^{\mathrm{T}}$ ($p \le r$) and $\bm{C}^{\mathrm{T}}\bm{C}$ ($p>r$) are larger and the computational time of the QD method is much shorter than those of the other calculation methods in the case of $1 \le p \le 20$ (the number of POD modes $r=10$ for the random sensor problem).

\subsection{PIV sensor problem}\label{subsec:PIV}

Real-time particle-image-velocimetry (PIV) measurement of the flowfield is required for the feedback control of a high-speed flowfield in laboratory experiments. The flowfield is calculated from the cross-correlation coefficient for each interrogation window of the particle image in the PIV measurement, but the number of windows that can be processed in a short time is limited because of the high computational costs of the calculation of cross-correlation coefficients. We have been developing a sparse-processing-particle-image-velocimetry(SPPIV)-measurement system~\cite{kanda2021feasibility}. The key point of this SPPIV-measurement system is that the amount of processing data is reduced and the flowfield is estimated by a limited number of sparsely located selected windows located, which allows the real-time PIV measurement of the flowfield. The development of an appropriate sensor selection method is required for the highly accurate SPPIV, and the framework of this study could prove useful for that task.

\begin{figure}[htbp]
    \centering
    \includegraphics[width=3in]{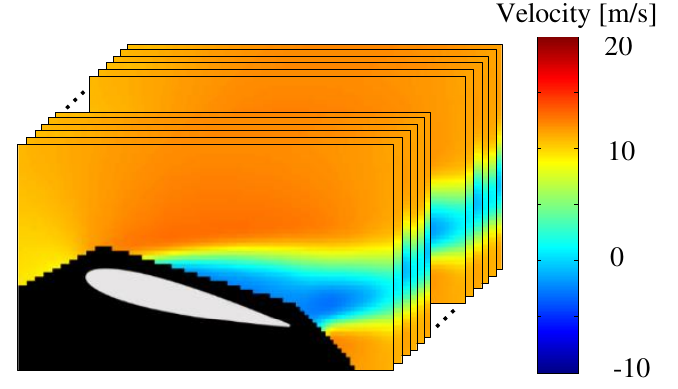}
    \caption{Dataset of PIV. Here, only the lateral components of the velocities are used for this problem.}
    \label{fig:PIV}
\end{figure}

A time-resolved dataset of the flowfield around an airfoil was acquired by a PIV measurement previously~\cite{nankai2019linear} (Fig. \ref{fig:PIV}) and is used for the reconstruction problem as a demonstration of a key technology used in SPPIV. 
Although the latent variable estimation and corresponding sensor selection methods might be better to be extended to the vector measurement\cite{saito2020data,kanda2021feasibility} owing to the nature of the PIV measurement, this problem is simplified to be one with the scalar measurement and is used for the verification of the proposed algorithm in this study.
Therefore, the lateral components of the velocities are only employed in this study, unlike the previous study in which the lateral and vertical components of the velocities are simultaneously treated\cite{saito2020data,kanda2021feasibility}. These PIV data with noise were obtained through real wind-tunnel experiments.

Here, the PIV data are briefly explained, and refer to \cite{nankai2019linear} for a detailed explanation. The wind tunnel testing was conducted in the Tohoku-University Basic Aerodynamic Research Wind Tunnel (T-BART) with a closed test section having a 300 mm $\times$ 300 mm cross-section. The airfoil of the test model had an NACA0015 profile, with the chord length and span width being 100 mm and 300 mm, respectively. The freestream velocity $U_\infty$ and the angle of attack for the airfoil $\alpha$ were set to be 10 m/s and 16 degrees, respectively. The chord Reynolds number was 6.4 $\times$ $10^4$. Time-resolved PIV measurement was conducted with a double-pulse laser (LDY-300PIV, Litron). The time between pulses, the sampling rate, the particle image resolution, and the total number of image pairs were \SI{100}{\micro\second}, 5000 Hz, 1024 $\times$ 1024 pixels, and $500$, respectively. The tracer particles were 50\% aqueous solution of glycerin with an estimated diameter of a few micrometers. The particle images were acquired by using the double-pulse laser and a high-speed camera (SA-X2, Photron), which were synchronized to each other.

\begin{figure}[htbp]
    \centering
    \includegraphics[width=3in]{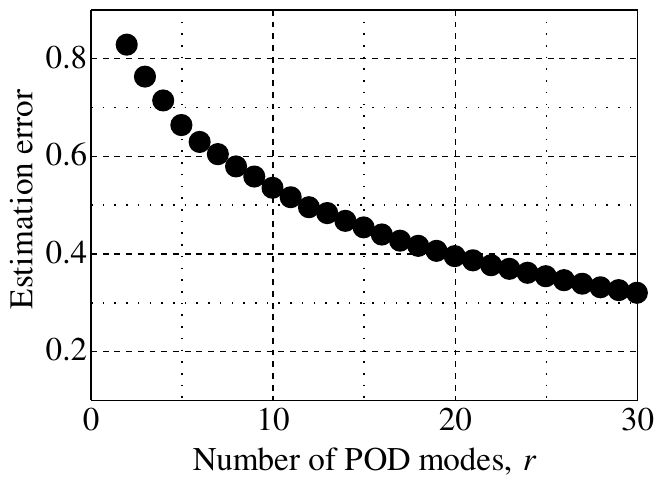}
    \caption{Estimation error against the number of POD modes for the PIV sensor problem. Estimation error here refers to the error from projecting the full state onto $r$ POD modes. Here, only the lateral components of the velocities are used for this problem.}
    \label{fig:truncation_PIV}
\end{figure}

\begin{figure}[htbp]
    \centering
    \includegraphics[width=3in]{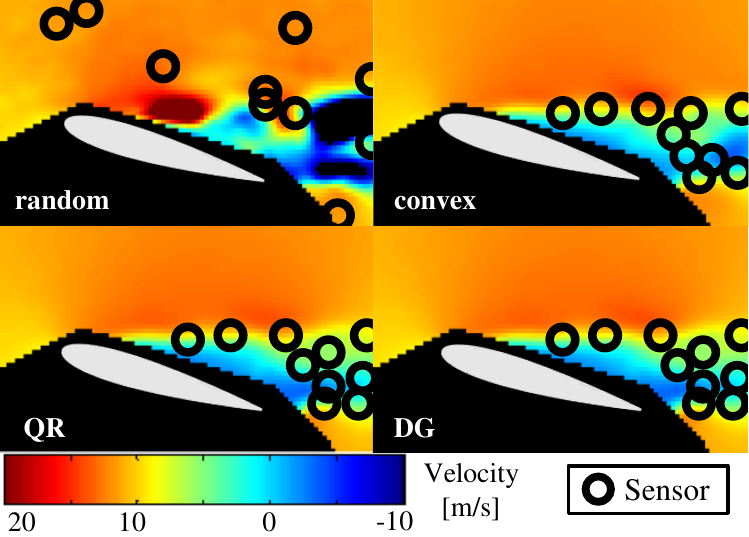}
    \caption{Sensor positions and reconstructed flowfield of a single snapshot from the sensors for $r=p=10$ in the PIV sensor problem. Here, only the lateral components of the velocities are used for this problem.}
    \label{fig:PIV_sensor_p=10}
\end{figure}

\begin{figure}[htbp]
    \centering
    \includegraphics[width=3in]{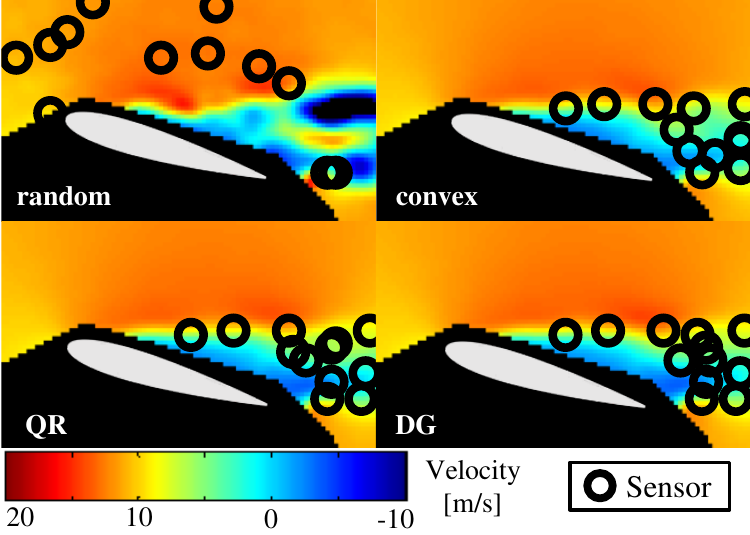}
    \caption{Sensor positions and reconstructed flowfield of a single snapshot for the sensors $r=10, p=12$ in the PIV sensor problem. Here, only the lateral components of the velocities are used for this problem.}
    \label{fig:PIV_sensor_p=12}
\end{figure}

The sparse sensor problem for the reconstruction of the lateral-velocity components of velocities measured by PIV is solved using the same computational environments in Table~\ref{table:Computational enviroment} as used in the random sensor problem.
Fig.~\ref{fig:truncation_PIV} shows the relationship between the estimation error and the number of POD modes, where the estimation error is defined as the ratio of the difference between the reconstructed data and the full observation data to the full observation. 
The estimation error $e$ is defined as follows:
\begin{eqnarray}
  e = \frac{1}{N}\sum_{j=1}^N\frac{\|\bm{x}_{j}-\hat{\bm{x}}_{j}\|^2_2}{\|\bm{x}_{j}\|^2_2}
    \label{eq:error}
\end{eqnarray}
where $\hat{\bm{x}}$ is the estimated data vector by using a sparse sensor measurement and competed as follows: $\hat{\bm{x}}=\bm{U}\bm{C}^{\dagger}\bm{y}$. Here, the subscript $j$~denotes the quantity of $j$th time step and $N$ represents the number of tested data. In the full observation, the estimated data vector is computed as follows: $\hat{\bm{x}}=\bm{U}_{r}\bm{U}_{r}^{\rm{T}}\bm{x}$.
The POD modes are truncated and an $r=10$ low-dimensional representation is obtained. Fig.~\ref{fig:truncation_PIV} shows that the minimum estimation error is 0.54 for $r=10$.
We consider the results of the sensor selection problem (the number of sensors $p=1$--$20$) solved by the methods: the random selection, convex approximation\cite{joshi2009sensor}, QR\cite{MANOHAR2018DATA}, DG, and QD methods.
Figs.~\ref{fig:PIV_sensor_p=10} and \ref{fig:PIV_sensor_p=12} show the sensor positions and the reconstructed flowfield of a single snapshot from the sensors selected by the random selection, convex approximation, QR, and QD methods in the cases of, respectively, $r=p=10$ and $r=10$, $p=12$. Note that all sensors of the QD method are located at the same positions as those of the DG method in all cases.
Fig.~\ref{fig:PIV_sensor_p=10} shows that the sensor position of the DG method is the same as that of the QR method in the case of $p \le r$, which confirms what is explained in Appendix \ref{Appendix:Equivalence of operations in the QR and present methods}. On the other hand, Fig.~\ref{fig:PIV_sensor_p=12} shows that the greedy added sensor positions of the QR and QD methods differ from each other in the case of $p > r$. 
The sensor positions of the QR and DG methods in Fig.~\ref{fig:PIV_sensor_p=12} are the same as the first 12 sensor positions of the QR and DG methods in the case of $p>12$ because the QR and DG methods are both greedy algorithms.
The quality of the sensors is evaluated by considering the estimation error. First, the estimation error in the case in which the training and validation data are the same as each other is discussed.

\begin{figure}[htbp]
    \centering
    \includegraphics[width=3in]{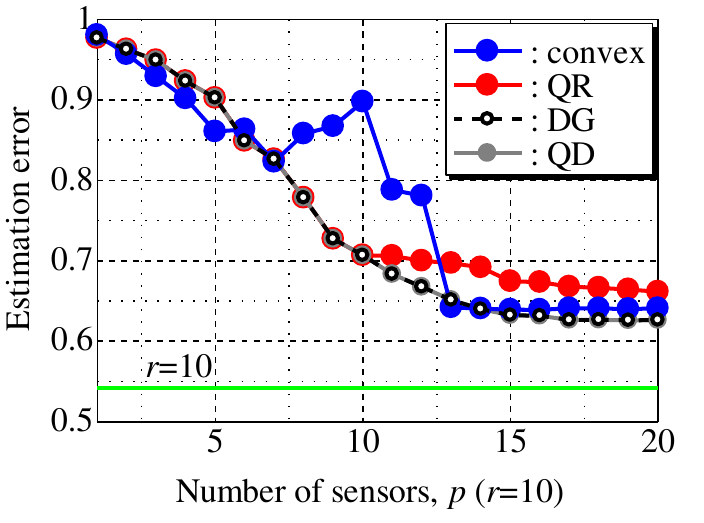}
    \caption{Estimation error against the number of sensors for the number of POD modes $r=10$ in the PIV sensor problem when the validation data are the same as the training data. Here, only the lateral components of the velocities are used for this problem. }
    \label{fig:error_PIV}
\end{figure}

Fig.~\ref{fig:error_PIV} shows the relationship between the estimation error and the number of sensors for the number of POD modes $r=10$ in the PIV problem. The estimation errors of the QR, DG, and QD methods decrease as the number of sensors increases, but the error of the convex approximation method increases in the case of $p=7$--$10$. Although the estimation error of the random selection is averaged over 1000 trials, all the values are outside the range shown in the vertical axis of the figure.
The estimation errors of the QR, DG, and QD methods coincide in the case of $p \le r$ because they give the same set of sensor locations. The estimation errors of the DG and QD methods remain the same as each other in the case of $p > r$, as in the random sensor problem. Although the estimation error of the QR method also decreases as the number of sensors increases, it provides inferior sensor positions compared to the DG or QD method.

\begin{figure}[htbp]
    \centering
    \includegraphics[width=3in]{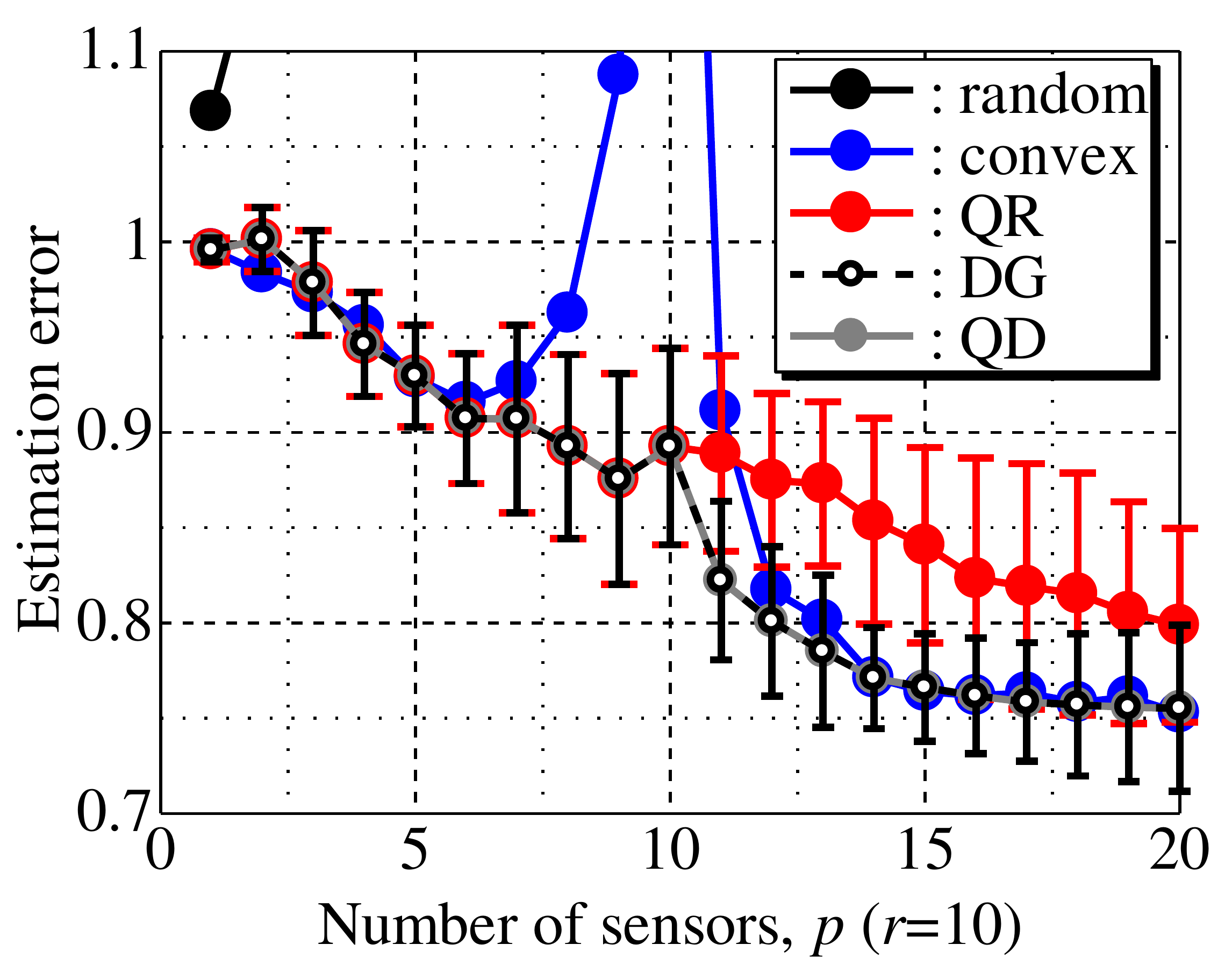}
    \caption{$K$-fold cross-validation results of estimation error against the number of sensors for the number of POD modes $r=10$ in the PIV sensor problem. Here, only the lateral components of the velocities are used for this problem.}
    \label{fig:error_PIV_CV}
\end{figure}

The estimation error for the case in which the training and validation data are different is discussed, which seems to be a more practical situation. A $K$-fold cross-validation ($K=5$) is conducted and the performances of algorithms are compared. The same PIV data set is used for the cross-validation. The $500$ snapshots are partitioned into five segments. Each of the five data segments is used as the validation data with the remaining data used for training and the estimation error is evaluated. 
Fig.~\ref{fig:error_PIV_CV} shows the results of this $K$-fold cross-validation ($K=5$) for a varying number of sensors $p$. The error bars represent the standard deviations of the five calculations, but the error bars of the random selection and convex approximation methods are omitted for easy understanding of the figure. Fig.~\ref{fig:error_PIV_CV} shows the same trend as that shown in Fig.~\ref{fig:error_PIV}, which generalizes that the characteristics of the sensors selected by each algorithm do not change when we look at more practical situations in the PIV problem.

\begin{figure}[htbp]
    \centering
    \includegraphics[width=3in]{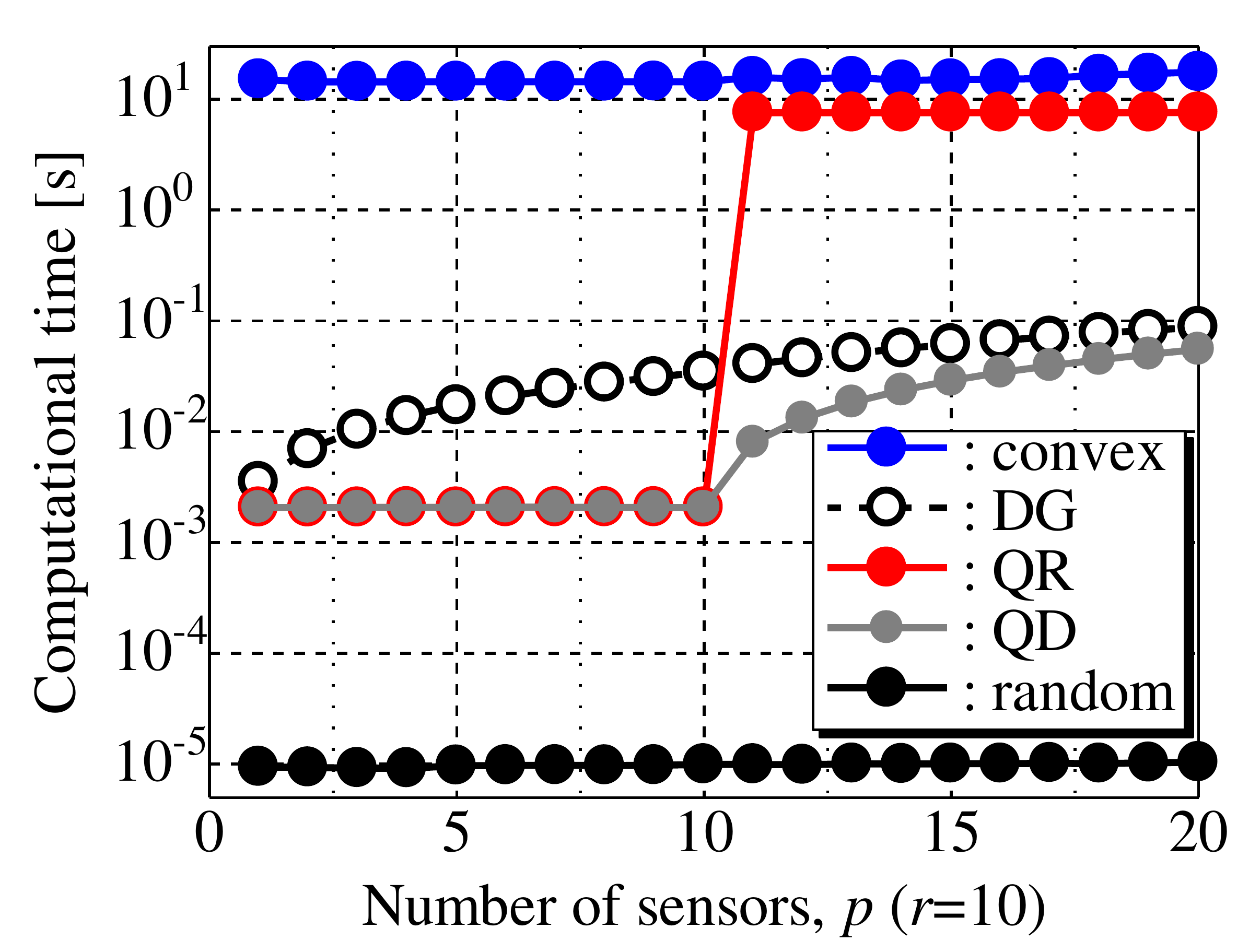}
    \caption{Computational time against the number of sensors for the number of POD modes $r=10$ in the PIV sensor problem. Here, only the lateral components of the velocities are used for this problem.}
    \label{fig:time_PIV}
\end{figure}

Fig.~\ref{fig:time_PIV} shows the computational time of each method for selecting sensors for the number of sensors $p=1$--$20$ in the PIV problem. Although the values of the random selection, DG, QR, and QD methods plotted in Fig.~\ref{fig:time_PIV} are averaged over 1000 calculations, those of the convex approximation method are averaged over 200 calculations due to the huge computational cost.
The trend in Fig.~\ref{fig:time_PIV} is the same as that in Fig. \ref{fig:Random_time}. These results show that the proposed method, referred to as the QD method herein, has the best properties in terms of shorter computational time and a more favorable sensor selection when the numbers of sensors and POD modes are $p=1$--$20$ and $r=10$, respectively.

\subsection{NOAA-SST sensor problem}\label{subsec:NOAA-SST}
\begin{figure}[htbp]
    \centering
    \includegraphics[width=3in]{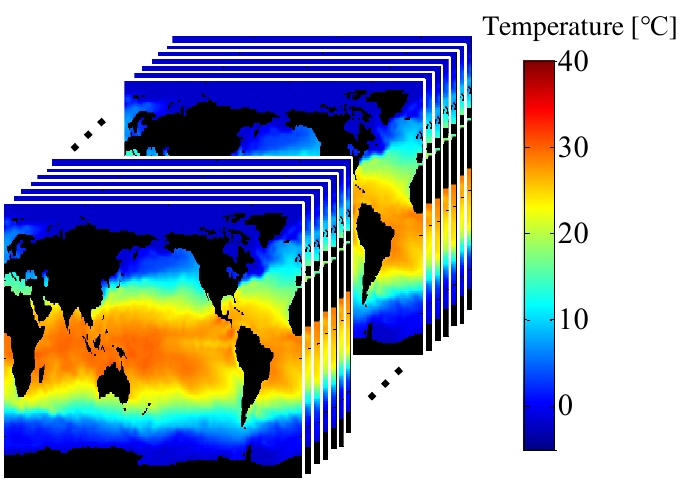}
    \caption{Dataset of NOAA-SST.}
    \label{fig:Data_set}
\end{figure}

\begin{figure}[htbp]
    \centering
   \includegraphics[width=3in]{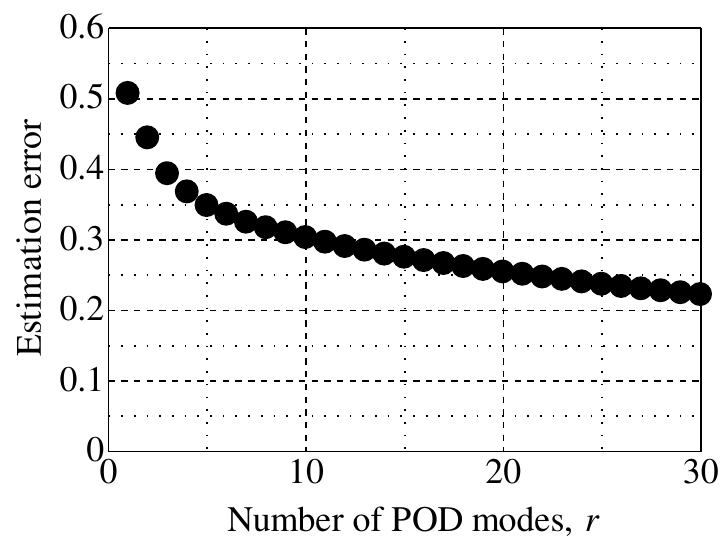}
    \caption{Estimation error against the number of POD modes for the NOAA-SST sensor problem.}
    \label{fig:EstimationError_POD}
\end{figure}

The data set that we finally adopt is the NOAA OISST (NOAA-SST) V2 mean sea surface temperature set, comprising weekly global sea surface temperature measurements in the years between 1990 and 2000. The data are publicly available online\cite{noaa}. There are a total of 520 snapshots on a 360 $\times$ 180 spatial grid (Fig.~\ref{fig:Data_set}). The sparse sensor problem for NOAA-SST is solved using the same computational environments in Table~\ref{table:Computational enviroment} as those used in the problems of the random sensor and PIV. 
Fig.~\ref{fig:EstimationError_POD} shows the relationship between the estimation error and the number of POD modes. In the case of NOAA-SST, the POD modes are truncated and the $r=10$ low-dimensional representation is obtained. The minimum estimation error is 0.30 for $r=10$ as Fig.~\ref{fig:EstimationError_POD} shows. We consider the results of the sensor selection problem (the number of sensors $p=10$--$20$) solved by each method: the random selection, convex approximation\cite{joshi2009sensor}, QR\cite{MANOHAR2018DATA}, DG, and QD methods. 

\begin{figure}[htbp]
    \centering
    \includegraphics[width=3in]{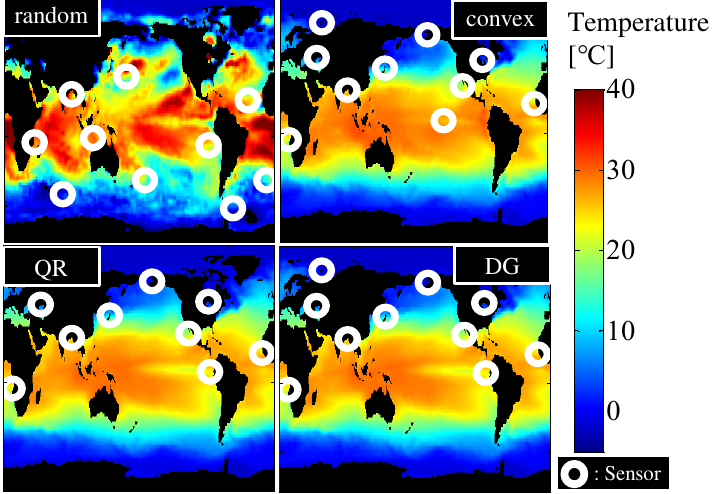}
    \caption {Sensor positions and reconstructed temperature field of a single snapshot from the sensors for $r=p=10$ in the NOAA-SST sensor problem.}
    \label{fig:Sensor_position_p=10}
\end{figure}

\begin{figure}[htbp]
    \centering
    \includegraphics[width=3in]{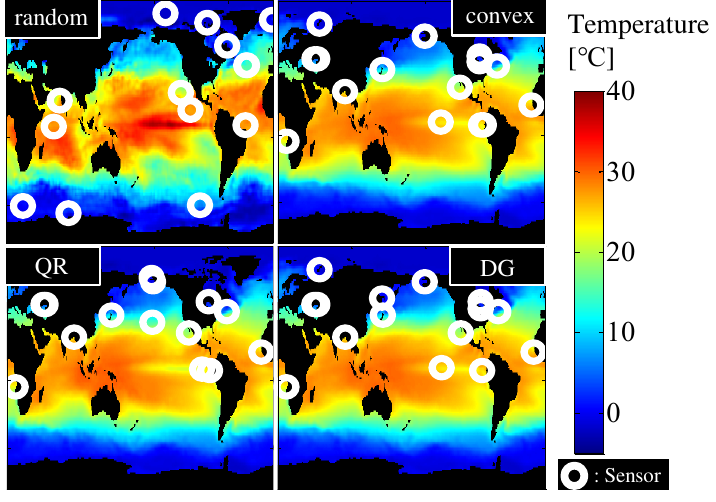}
    \caption {Sensor positions and reconstructed temperature field of a single snapshot from the sensors for $r=10, p=15$ in the NOAA-SST sensor problem.}
    \label{fig:Sensor_position_p=15}
\end{figure}

Figs.~\ref{fig:Sensor_position_p=10} and \ref{fig:Sensor_position_p=15} show the sensor positions selected by the random selection, convex approximation, QR, and QD methods in the cases of, respectively, $r=p=10$ and $r=10$, $p=15$. All sensors of the QD method are located at the same positions as those of the DG method in all cases, as in the problems of the random sensor and PIV.

\begin{figure}[htbp]
    \centering
    \includegraphics[width=3in]{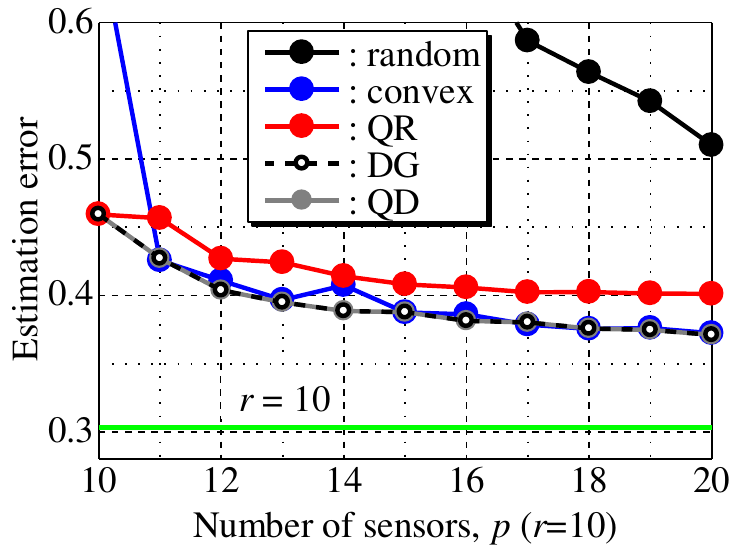}
    \caption {Estimation error against the number of sensors for the number of POD modes $r=10$ in the NOAA-SST sensor problem when the validation data are the same as the training data.}
    \label{fig:EstimationError}
\end{figure}

The quality of the sensors is evaluated by considering the estimation error. First, the estimation error in the case in which the training and validation data are the same as each other is discussed. 
Fig.~\ref{fig:EstimationError} shows the relationship between the estimation error and the number of sensors for the number of POD modes $r=10$ in the NOAA-SST problem. All estimation errors decrease as the number of sensors increases. The estimation errors of the QR, DG, and QD methods coincide in the case of $p=10$ because they give the same set of sensor locations. The estimation errors of the DG and QD methods remain the same as each other in the case of $p>r$, as in the random sensor problem. Although the estimation error of the QR method also decreases as the number of sensors increases, it provides inferior sensor positions compared to the DG or QD method. 

\begin{figure}[htbp]
    \centering
    \includegraphics[width=3in]{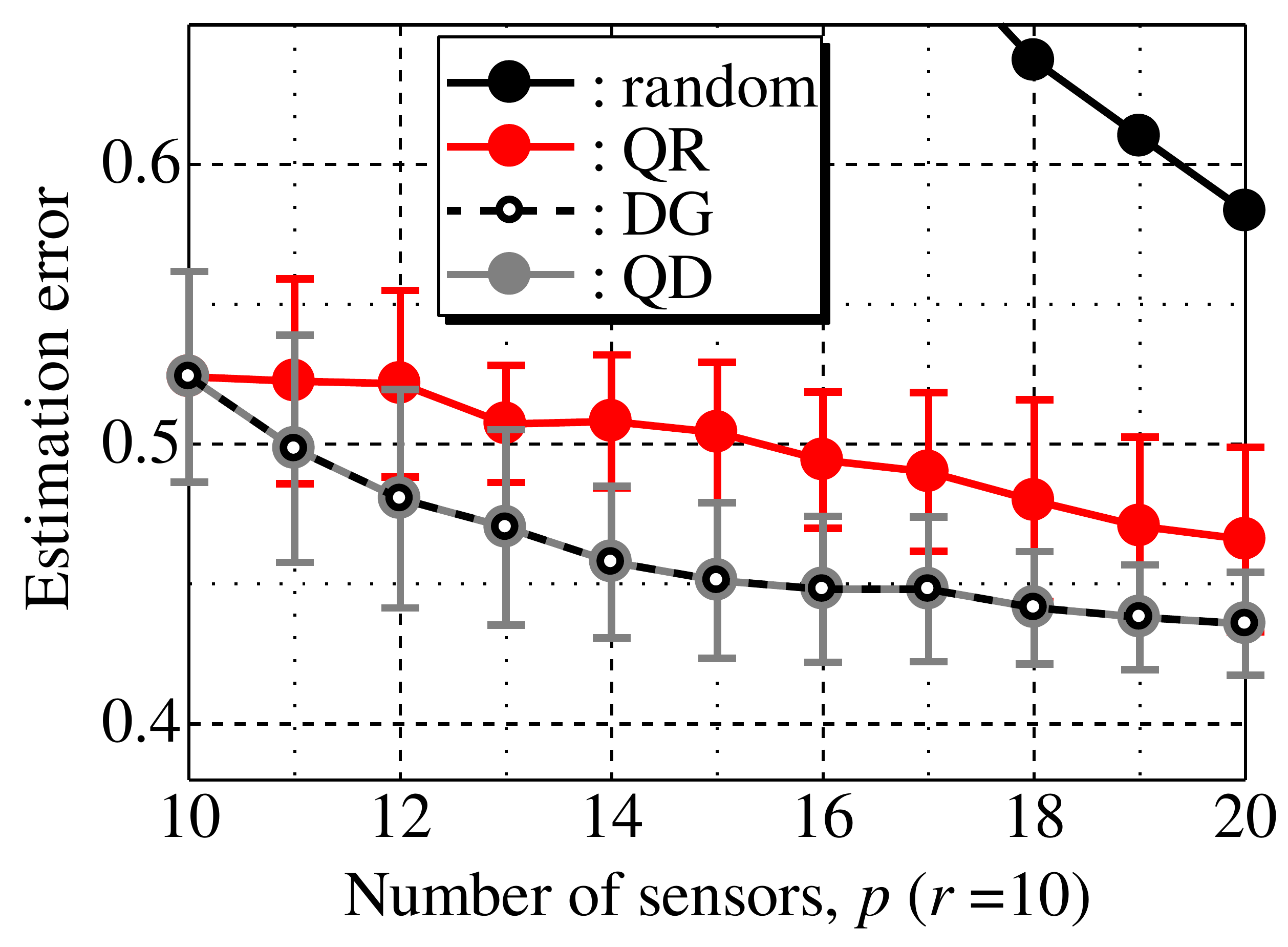}
    \caption{$K$-fold cross-validation results of estimation error against the number of sensors for the number of POD modes $r=10$ in the NOAA-SST sensor problem.}
    \label{fig:CrossValication}
\end{figure}

Next, the estimation errors for the case in which the training and validation data are different are considered, which seems to be a more practical situation. A $K$-fold cross-validation ($K=5$) is conducted and the performances of algorithms are compared. The same NOAA-SST data set is used for the cross-validation. The $520$ snapshots are partitioned into five segments. Each of the five data segments is used as the validation data with the remaining data used for training and the estimation error is evaluated. Fig.~\ref{fig:CrossValication} shows the results of this $K$-fold cross-validation ($K=5$) for a varying number of sensors $p$. In the figure, error bars represent the standard deviations of the five calculations. The convex method is not included, because its computational time is very long in the NOAA-SST problem. Fig.~\ref{fig:CrossValication} shows the same trend as that shown in Fig.~\ref{fig:EstimationError}, which generalizes that the characteristics of the sensors selected by each algorithm do not change when more practical situations are considered.
Therefore, both the DG and QD methods in the case of $p>r$ are more accurate sensor selection methods than the QR method, in which the QR decomposition of $\bm{UU}^{\mathrm{T}}$ is performed.

\begin{figure}[htbp]
    \centering
    \includegraphics[width=3in]{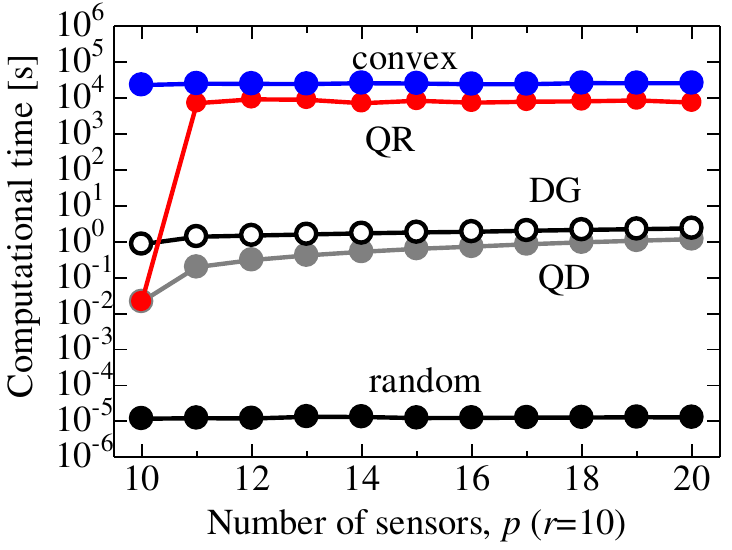}
    \caption {Computational time against the number of sensors for the number of POD modes $r=10$ in the NOAA-SST sensor problem.}
    \label{fig:CalculationTime}
\end{figure}

Fig.~\ref{fig:CalculationTime} shows the computational time of each method for selecting sensors for the number of sensors $p=10$--$20$. 
The computational time of the QR method has a large difference at $p=11$ because the dimension of the matrix processed by the QR decomposition increases between the computation for $\bm{U}\in \mathbb{R}^{n{\times}r}$ and that for $\bm{UU}^{\mathrm{T}}\in \mathbb{R}^{n{\times}n}$. This trend is the same as that of the problems of the random sensor and PIV. 
On the other hand, the computational time of the DG method gradually increases as the number of sensors increases because each sensor is chosen using the step-by-step maximization of the determinant of $\bm{C}_{k}^{\mathrm{T}}\bm{C}_{k}$ in the case of $p>r$.
Therefore, the complexities of the QR and DG/QD methods are ${\mathcal O}\left(n^3 \right)$ and ${\mathcal O}\left(pnr^2 \right)$ in the case of $p>r$, respectively.
Hence, for the high-dimensional data such as the NOAA-SST sensor problem, the computational time of the QD method is much shorter than that of the QR method in the case of $p>r$. 
Although the trend in Fig.~\ref{fig:CalculationTime} is the same as that in Figs.~\ref{fig:Random_time} and \ref{fig:time_PIV}, note that the computational time of the convex approximation method is approximately three hours, whereas that of the QD method, which gives a lower estimation error, is just a few seconds. These results show that the proposed method, referred to as the QD method herein, has the best properties in terms of shorter computational time and a more favorable sensor selection when the numbers of sensors and POD modes are $p=10$--$20$ and $r=10$, respectively. Based on the proposed determinant-based greedy algorithm, we are currently working on implementations that take noise into consideration\cite{yamada2021fast} and reevaluate the objective functions\cite{nakai2021effect}.

\section{Conclusions}
Optimal sensor placement is an important challenge in the design, prediction, estimation, and control of high-dimensional systems. In this study, the sparse sensor placement problem is considered for least-squares estimation. First, the objective function of the problem is redefined for the maximization of the determinant of the matrix appearing in pseudo-inverse matrix operations, leading to the maximization of the corresponding confidence intervals. The procedure for the maximization of the determinant of the corresponding matrix is proved to be mathematically equivalent to that of the previous QR method when the number of sensors is less than that of state variables. On the other hand, for the case that the number of sensors is greater than that of state variables, we developed a new algorithm then, a unified formulation is derived, and the lower bound of the objective function given by this algorithm is shown using its monotone submodularity. In the proposed algorithm, optimal sensors are obtained by the QR method until the number of sensors is equal to that of state variables, beyond which, new sensors are calculated by the proposed determinant-based greedy method which is accelerated by both a determinant formula and a matrix inversion lemma. The effectiveness of this algorithm on datasets related to the flowfield around an airfoil and the global climate is demonstrated in comparison to other algorithms. The computational time of the proposed extended determinant-based greedy algorithm is shown to be smaller than that of other methods with one of the smallest estimation errors. One weakness of the proposed algorithm is its robustness to noise, but it has been addressed and improved in \cite{yamada2021fast}. The further improvement of the greedy method will be the subject of challenging future research.
We have been developing a sparse-processing-particle-image-velocimetry(SPPIV)-measurement system~\cite{kanda2021feasibility}. In the SPPIV-measurement system, the amount of processing data is reduced and the flowfield is estimated by a limited number of sparsely located interrogation windows, leading to the real-time PIV measurement of the flowfield. The development of an appropriate sensor selection method is required for the highly accurate SPPIV, and the proposed algorithm combined with the vector extension\cite{saito2020data} could be useful for that task.
In addition to the SPPIV-measurement system, the proposed method, which can significantly reduce the computational time of selecting the sensor position compared to the previous studies, is suitable for application to particularly high-dimensional data such as combustion data.
\appendices

\section{Summary of previous greedy algorithm and QR decomposition in the oversampling case}
\begin{algorithm}
\label{alg1}                          
\caption{QR algorithm for sparse sensor placement}\label{alg:previous}
\begin{algorithmic}
\STATE Set sensor-candidate matrix $\bm{U}$.
\STATE Set number of sensors $p \geq r$.
\IF{$p=r$}
\STATE$\bm{V}=\bm{U}$
\ELSIF{$p>r$}
\STATE $\bm{V}=\bm{UU}^{\mathrm{T}}$
\ENDIF
\FOR{ $k =1, \dots, p$ }
      \STATE $\bm{v}_{i}=[\begin{array}{cccc}V_{i1} &V_{i2} &\dots &V_{ir}\end{array}]$
      \STATE $ i  \leftarrow \argmax_{i\, \in\, \mathcal{S}\, \backslash\, \mathcal{S}_{k}} \|\bm{v}_i \|^2_2$
      \STATE $\bm{w}_k \leftarrow \bm{v}_i $
      \STATE $\bm{V} \leftarrow \bm{V}-\bm{V}\bm{w}_k^{\mathrm{T}}\bm{w}_k/\|\bm{w}_k\|^2_2$
      \STATE $H_{k,i} = 1$
\ENDFOR
\end{algorithmic}
\end{algorithm}

\begin{figure}[htbp]
    \centering
    \includegraphics[width=3.3in]{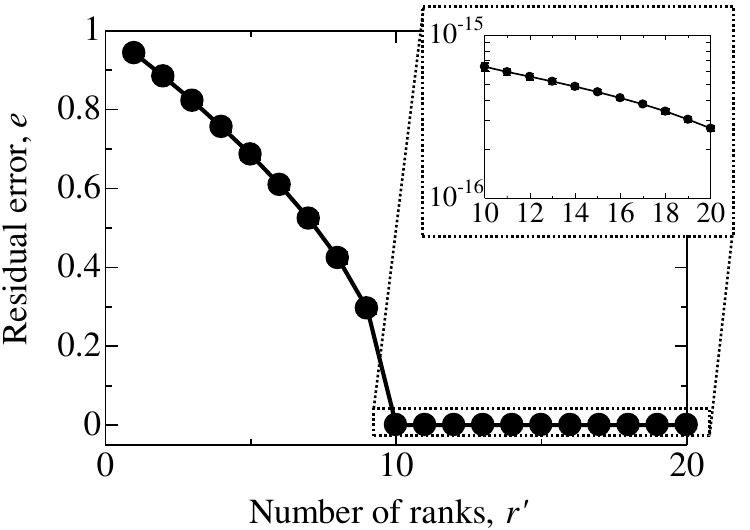}
    \caption{Residual error against the number of ranks at the implementation of QR decomposition for $\bm{UU}^{\mathrm{T}}$.}
    \label{fig:QR oversampling}
\end{figure}

\label{Appendix:Summary of previous greedy algorithm and QR decomposition in the oversampling case}
The QR algorithm is presented as in Alg. \ref{alg:previous}. The optimization is to maximize the determinant of the matrix $\bm{C}$ to stably solve for the $\bm{x}$ vector in the case of $p=r$. On the other hand, the validity of the algorithm in the oversampling case ($p>r$) was not well clarified in\cite{MANOHAR2018DATA}.
Here, the QR method in oversampling is considered. The random sensor-candidate  matrices, $\bm{U}\in \mathbb{R}^{1000{\times}r}$, were set, where the component of the matrices is given by the Gaussian distribution $\mathcal{N}(0,1)$ and $r=10$. The QR decomposition is carried out as follows:
\begin{equation}
\bm{Q}\bm{R}=\bm{UU}^{\mathrm{T}},
\end{equation}
where $\bm{Q}\in \mathbb{R}^{1000{\times}1000}$ is an orthogonal matrix (its columns are orthogonal unit vectors) and $\bm{R}\in \mathbb{R}^{1000{\times}r}$ is an upper triangular matrix.
Fig. \ref{fig:QR oversampling} shows the residual error, $e$, against the number of ranks, $r^{\prime}$. Here, the residual error expresses the following equation:
\begin{equation} 
e=\frac{\left|\left|\bm{Q}\bm{R}-\bm{Q}^{\prime}\bm{R}^{\prime}\right|\right|_{F}}{\left|\left|\bm{Q}\bm{R}\right|\right|_{F}},
\end{equation}
where $\bm{Q}^{\prime}$ and $\bm{R}^{\prime}$ denote the matrices that consist of the first $r^{\prime}$ columns of $\bm{Q}$ and the first $r^{\prime}$ rows of $\bm{R}$, respectively, and the subscript $F$ expresses the Frobenius norm.
Fig. \ref{fig:QR oversampling} shows that the reconstruction with the QR decomposition of $r^{\prime}=r$ is almost the same as the original matrix $\bm{UU}^{\mathrm{T}}$.  This is because the rank of $\bm{UU}^{\mathrm{T}}$ is mathematically $r$ and the residual error of $r^{\prime} \ge r=10$ is the computationally generated machine epsilon $\left(O\left(10^{-16}\right)\right)$ in double-precision decimals. Therefore, the QR decomposition works only for the residual of the machine epsilon level in the oversampling case. However, the QR method selects worse and better sensor positions than the present method and the random method, respectively, as explained in Appendix \ref{Appendix:Comparison to previous studies in oversampling case}. This is because the QR method selects the sensor positions where the residual of the machine epsilon level is large (the sensor position where the value of $\bm{U}^{\mathrm{T}}\bm{U}$ is possibly large). As a result, it accidentally selects somehow reasonable sensor positions rather than randomized positions, but it is not optimized for the sensor selection problem in the oversampling case.

\section{Equivalence of operations in the QR and present methods for $k \le r$}
\label{Appendix:Equivalence of operations in the QR and present methods}
The proof of Theorem \ref{theorem:QR=DG_in_the_k_less_than_p} is given as follows.
\begin{proof}
We start by deriving a mathematical expression of the QR method.
The first step is to find a row vector of $\bm{U}$ with the largest norm.
Suppose that $\bm{w}_{1} \in \mathbb{R}^{1 \times r }$ is such a vector and that it is in the $i_{1}$th row of $\bm{U}$.
Namely, $\bm{w}_{1} = \bm{u}_{i_{1}}$.
In the next step, we subtract components parallel to $\bm{w}_{1}$ from every row vector of $\bm{U}$.
This operation is represented as
\begin{equation}
 \bm{U} 
 \left(
   \bm{I} - \frac{\bm{w}_{1}^{\mathrm{T}} \bm{w}_{1}^{} }{ \| \bm{w}_{1} \|^{2} } 
 \right)
 =: \bm{U} \bm{P}_{1}^{\mathrm{QR}}.
\end{equation}
Note that the matrix $\bm{P}_{1}^{\mathrm{QR}} \in \mathbb{R}^{r \times r} $ satisfies $\bm{P}_{1}^{\mathrm{QR}} = \left(\bm{P}_{1}^{\mathrm{QR}}\right)^{ \mathrm{T}}$ and $(\bm{P}_{1}^{\mathrm{QR}})^{2} = \bm{P}_{1}^{\mathrm{QR}}  $.
Hence, $\bm{P}_{1}^{\mathrm{QR}} $ is a symmetric projection matrix.
We then seek a row vector $\bm{w}_{2} \in \mathbb{R}^{1 \times r}$ of $\bm{U} \bm{P}_{1}^{\mathrm{QR}}$ with the largest norm.
Let us suppose that $\bm{w}_{2} = \bm{v}_{i_{2}}$, which is the $i_{2}$th row vector of $\bm{U} \bm{P}_{1}^{\mathrm{QR}}$.
Clearly, $\bm{w}_{2}$ satisfies $\bm{w}_{2} = \bm{u}_{i_{2}}\bm{P}_{1}^{\mathrm{QR}} $ and $\bm{w}_{2}^{} \bm{w}_{1}^{\mathrm{T}} = 0 $.
As in the previous steps, we consider the matrix
\begin{equation}
 \bm{U} \bm{P}_{1}^{\mathrm{QR}} 
 \left(
   \bm{I} - \frac{\bm{w}_{2}^{\mathrm{T}} \bm{w}_{2}^{} }{ \| \bm{w}_{2} \|^{2} } 
 \right) 
 =: \bm{U} \bm{P}_{2}^{\mathrm{QR}}
\end{equation}
and find a row vector of $\bm{U} \bm{P}_{2}^{\mathrm{QR}}$ with the largest norm.
Since $\bm{w}_{2} \bm{w}_{1}^{\top} = 0$, we have $\bm{w}_{2}^{\mathrm{T}}\bm{w}_{2}\bm{P}_{1}^{\mathrm{QR}}=\bm{P}_{1}^{\mathrm{QR}}\bm{w}_{2}^{\mathrm{T}}\bm{w}_{2}$. This yields
\begin{align}
 {\bm{P}_{2}^{\mathrm{QR}}}^{\mathrm{T}}
 &=
 \left(
   \bm{I} - \frac{\bm{w}_{2}^{\mathrm{T}} \bm{w}_{2}^{} }{ \| \bm{w}_{2} \|^{2} } 
 \right)\bm{P}_{1}^{\mathrm{QR}}
 =
 \bm{P}_{1}^{\mathrm{QR}}\left(
   \bm{I} - \frac{\bm{w}_{2}^{\mathrm{T}} \bm{w}_{2}^{} }{ \| \bm{w}_{2} \|^{2} } 
 \right) 
 =\bm{P}_{2}^{\mathrm{QR}}, \notag \\
 {\bm{P}_{2}^{\mathrm{QR}}}^{2}
 &=
 \bm{P}_{1}^{\mathrm{QR}}\left(
   \bm{I} - \frac{\bm{w}_{2}^{\mathrm{T}} \bm{w}_{2}^{} }{ \| \bm{w}_{2} \|^{2} } 
 \right)
 \bm{P}_{1}^{\mathrm{QR}}\left(
   \bm{I} - \frac{\bm{w}_{2}^{\mathrm{T}} \bm{w}_{2}^{} }{ \| \bm{w}_{2} \|^{2} } 
 \right) \notag \\
 &= \bm{P}_{1}^{\mathrm{QR}}\left(
   \bm{I} - \frac{\bm{w}_{2}^{\mathrm{T}} \bm{w}_{2}^{} }{ \| \bm{w}_{2} \|^{2} } 
 \right)=
 \bm{P}_{2}^{\mathrm{QR}}.
\end{align}
Thus, $\bm{P}_{2}^{\mathrm{QR}} $ is also a symmetric projection.
In this way, we can define a sequence of symmetric projection matrices $\{ \bm{P}_{k}^{\mathrm{QR}} \}_{k = 0}^{r}$ by
\begin{equation}
 \bm{P}_{k}^{\mathrm{QR}}
 := \bm{P}_{k - 1}^{\mathrm{QR}}
 \left(
   \bm{I} - \frac{\bm{w}_{k}^{\mathrm{T}} \bm{w}_{k}^{} }{ \| \bm{w}_{k} \|^{2} } 
 \right), \
 k = 1, 2, \dots, r,
\end{equation}
where $\bm{P}_{0}^{\mathrm{QR}} = \bm{I}$ and $\bm{w}_{k} \in \mathbb{R}^{1 \times r}$ is the vector selected in the $k$th step of the QR method.
Suppose that $\bm{w}_{k}$ is the $i_{k}$th row vector, $\bm{v}_{i_{k}}$, of $\bm{U} \bm{P}_{k -1}^{\mathrm{QR}}$.
Then, we have $\bm{w}_{k} = \bm{v}_{i_{k}} = \bm{u}_{i_{k}} \bm{P}_{k-1}^{\mathrm{QR}}$.

Now, notice that the squared 2-norm of $\bm{w}_{k} = \bm{v}_{i_{k}}$ is given by $\| \bm{w}_{k} \|^{2} = \bm{u}_{i_{k}}^{} \bm{P}_{k - 1}^{\mathrm{QR}}  \bm{u}_{i_{k}}^{\mathrm{T}}$.
Hence, $\bm{w}_{k}$ can be characterized as
\begin{equation}
 \begin{aligned}
 \bm{w}_{k} 
&= \bm{v}_{i_{k}} 
 =\bm{u}_{i_{k}} \bm{P}_{k - 1}^{\mathrm{QR}}, \\ 
 i_{k}
&= \argmax_{i}
   \bm{u}_{i}^{} \bm{P}_{k - 1}^{\mathrm{QR}}  \bm{u}_{i}^{\mathrm{T}}.
 \end{aligned}
\end{equation}
In the DG method explained in Alg.~\ref{alg:dgr_k_le_r}, we choose a row vector of $\bm{U}$ that  
maximizes a quadratic form involving matrix $\bm{C}$.
This fact motivates us to relate this matrix to $\bm{P}_{k}^{\mathrm{QR}}$.

In the remainder of the proof, we show that $\bm{P}_{k}^{\mathrm{QR}} = \bm{I} - \bm{C}_{k}^{\mathrm{T}} \left(\bm{C}_{k}^{}\bm{C}_{k}^{\mathrm{T}}\right)^{-1}\bm{C}_{k}^{}$, where 
\begin{equation}
 \bm{C}_{k}
 = \left[
      \begin{array}{@{\,}cccc@{\,}}
         \bm{u}_{i_{1}}^{\mathrm{T}} &
         \bm{u}_{i_{2}}^{\mathrm{T}} & \dots &
         \bm{u}_{i_{k}}^{\mathrm{T}} 
      \end{array}
   \right]^{\mathrm{T}}.
\end{equation}
Since $\bm{w}_{1} = \bm{u}_{i_{1}}$, we have $\bm{C}_{1} = \bm{u}_{i_{1}}$.
It immediately follows that
\begin{equation}
   \bm{I} - \bm{C}_{1}^{\mathrm{T}} (\bm{C}_{1}^{}\bm{C}_{1}^{\mathrm{T}})^{-1}\bm{C}_{1}^{}
 = \bm{I} 
   - \frac{\bm{u}_{i_{1}}^{\mathrm{T}} \bm{u}_{i_{1}}^{}}{ \bm{u}_{i_{1}}^{} \bm{u}_{i_{1}}^{\mathrm{T}} }
 = \bm{I} 
   - \frac{\bm{w}_{1}^{\mathrm{T}} \bm{w}_{1}^{}}{ \| \bm{w}_{1} \|^{2} }
 = \bm{P}_{1}^{\mathrm{Q R}}.
\end{equation}
To prove our claim by induction, we assume that 
$\bm{P}_{k-1}^{\mathrm{QR}} = \bm{I} - \bm{C}_{k-1}^{\mathrm{T}} \left(\bm{C}_{k-1}^{}\bm{C}_{k-1}^{\mathrm{T}}\right)^{-1}\bm{C}_{k-1}^{}$.
Under this assumption, we represent the matrix
$\bm{I} - \bm{C}_{k}^{\mathrm{T}} \left(\bm{C}_{k}^{}\bm{C}_{k}^{\mathrm{T}}\right)^{-1}\bm{C}_{k}^{}$
in terms of $P_{k-1}^{\mathrm{QR}}$.
The second term can be written as
\begin{align}
&\bm{C}_{k}^{\mathrm{T}} \left(\bm{C}_{k}^{}\bm{C}_{k}^{\mathrm{T}}\right)^{-1}\bm{C}_{k}^{}
 \notag
 \\
&= \left[
     \begin{array}{@{\,}cc@{\,}}
       \bm{C}_{k - 1}^{\mathrm{T}} & \bm{u}_{i_{k}}^{\mathrm{T}}
     \end{array}
   \right]
   \left[
     \begin{array}{@{\,}cc@{\,}}
       \bm{C}_{k - 1}^{} \bm{C}_{k - 1}^{\mathrm{T}} &
       \bm{C}_{k - 1}^{} \bm{u}_{i_{k}}^{\mathrm{T}} \\
       \bm{u}_{i_{k}}^{} \bm{C}_{k - 1}^{\mathrm{T}} &
       \bm{u}_{i_{k}}^{} \bm{u}_{i_{k}}^{\mathrm{T}}
     \end{array}
   \right]^{-1}
   \left[
     \begin{array}{@{\,}c@{\,}}
       \bm{C}_{k - 1} \\ \bm{u}_{i_{k}}
     \end{array}
   \right].
 \label{eq:CkT_inv_CkCkT_Ck}
\end{align}
Let $\delta \in \mathbb{R}$ be the Schur complement of the block square matrix in the middle of the right-hand side
of \eqref{eq:CkT_inv_CkCkT_Ck}.
That is,
\begin{align}
 \delta
&=   \bm{u}_{i_{k}}^{} \bm{u}_{i_{k}}^{\mathrm{T}}
  -  \bm{u}_{i_{k}}^{} \bm{C}_{k - 1}^{\mathrm{T}}
    (\bm{C}_{k - 1}^{} \bm{C}_{k - 1}^{\mathrm{T}} )^{-1}
     \bm{C}_{k - 1}^{} \bm{u}_{i_{k}}^{\mathrm{T}}
 \notag \\
&=   \bm{u}_{i_{k}}^{} P_{k - 1}^{\mathrm{QR}}
     \bm{u}_{i_{k}}^{\mathrm{T}}
 \notag \\
&= \| \bm{w}_{k}^{} \|^{2}.
\end{align}
Then, by a property of the inverses of block matrices,
\begin{align}
&  \left[
     \begin{array}{@{\,}cc@{\,}}
       \bm{C}_{k - 1}^{} \bm{C}_{k - 1}^{\mathrm{T}} &
       \bm{C}_{k - 1}^{} \bm{u}_{i_{k}}^{\mathrm{T}} \\
       \bm{u}_{i_{k}}^{} \bm{C}_{k - 1}^{\mathrm{T}} &
       \bm{u}_{i_{k}}^{} \bm{u}_{i_{k}}^{\mathrm{T}}
     \end{array}
   \right]^{-1} 
 \notag \\
&= \left[
     \begin{array}{@{\,}cc@{\,}}
        \bm{Q}_{k - 1}^{-1} & 0 \\
        0 & 0
     \end{array}
   \right]  
  +\frac{1}{\delta}
   \left[
     \begin{array}{@{\,}c@{\,}}
          \bm{Q}_{k-1}^{-1} 
          \bm{C}_{k - 1}^{} \bm{u}_{i_{k}}^{\mathrm{T}}
    \\ - 1 
     \end{array}
   \right]
   \left[
     \begin{array}{@{\,}cc@{\,}}
          \bm{u}_{i_{k}}^{} \bm{C}_{k - 1}^{\mathrm{T}}
          \bm{Q}_{k-1}^{-1} 
        & -1
     \end{array}
   \right],
\end{align}
where $\bm{Q}_{k - 1} := \bm{C}_{k - 1}^{} \bm{C}_{k - 1}^{\mathrm{T}}$.
Substituting this result into \eqref{eq:CkT_inv_CkCkT_Ck} gives
\begin{align}
& \bm{C}_{k}^{\mathrm{T}}
 \left(\bm{C}_{k}^{}\bm{C}_{k}^{\mathrm{T}}\right)^{-1}\bm{C}_{k}^{} 
 = \bm{I} - \bm{P}_{k-1}^{\mathrm{QR}}
   + \frac{1}{\delta}
     \bm{P}_{k -1}^{\mathrm{QR}} \bm{u}_{i_{k}}^{\mathrm{T}} \bm{u}_{i_{k}}^{}
     \bm{P}_{k -1}^{\mathrm{QR}}.
\end{align}
This relation readily shows that
\begin{align}
  \bm{I} - \bm{C}_{k}^{\mathrm{T}}
 \left(\bm{C}_{k}^{}\bm{C}_{k}^{\mathrm{T}}\right)^{-1}\bm{C}_{k}^{} 
& =   \bm{P}_{k -1}^{\mathrm{QR}} 
    - \frac{1}{\delta} 
      \bm{P}_{k -1}^{\mathrm{QR}} \bm{u}_{i_{k}}^{\mathrm{T}} \bm{u}_{i_{k}}^{}
      \bm{P}_{k -1}^{\mathrm{QR}}
      \notag \\
& =   \bm{P}_{k -1}^{\mathrm{QR}} 
    - \frac{1}{\delta} 
     \left(\bm{P}_{k -1}^{\mathrm{QR}}\right)^{2} \bm{u}_{i_{k}}^{\mathrm{T}} \bm{u}_{i_{k}}^{}
      \bm{P}_{k -1}^{\mathrm{QR}}
      \notag \\
& =   \bm{P}_{k -1}^{\mathrm{QR}} 
      \left(
        \bm{I}
        - \frac{1}{\delta}
          \bm{P}_{k -1}^{\mathrm{QR}} \bm{u}_{i_{k}}^{\mathrm{T}} \bm{u}_{i_{k}}^{}
          \bm{P}_{k -1}^{\mathrm{QR}}
      \right)
      \notag \\
& =   \bm{P}_{k -1}^{\mathrm{QR}} 
      \left(
        \bm{I}
        - \frac{\bm{w}_{k}^{\mathrm{T}} \bm{w}_{k} }{\| \bm{w}_{k} \|^{2}}
      \right)
  \notag \\
& = \bm{P}_{k}^{\mathrm{QR}}.
\end{align}
Thus, mathematical induction guarantees that
$\bm{P}_{k}^{\mathrm{QR}} = \bm{I} - \bm{C}_{k}^{\mathrm{T}}\left(\bm{C}_{k}^{}\bm{C}_{k}^{\mathrm{T}}\right)^{-1}\bm{C}_{k}^{}$
for any $k \in \{ 1, 2, \dots, r\}$.
This relation means that, for each $k \in \{1 ,2, \dots, r\}$,
$\bm{w}_{k}$ is obtained by maximizing the quadratic form 
\begin{equation}
 \bm{u}_{i}^{} 
 \biggl(
    \bm{I}
  - \bm{C}_{k-1}^{\mathrm{T}}(\bm{C}_{k-1}^{}\bm{C}_{k-1}^{\mathrm{T}})^{-1}\bm{C}_{k-1}^{}
 \bigg) \bm{u}_{i}^{\mathrm{T}}
\end{equation}
with respect to $i$.
This is what is described in Alg.~\ref{alg:dgr_k_le_r}.
Hence, the QR method coincides with our DG method for $k \le r$.
\end{proof}

\section{Comparison to previous studies in oversampling case}
\label{Appendix:Comparison to previous studies in oversampling case}

Previous studies have introduced several methods for the sensor (sampling point) selection problem under oversampling ($p>r$). In the QR method\cite{MANOHAR2018DATA}, sensor selection is conducted by maximizing the determinant of the matrix $\bm{CC}^{\mathrm{T}}$ in the case of $p>r$. The calculation algorithm for the condition $p>r$ has already been proposed in\cite{MANOHAR2018DATA} (which corresponds to the use of $\bm{V}=\bm{UU}^{\mathrm{T}}$), but the validity of the algorithm was not well clarified in\cite{MANOHAR2018DATA}. The ``GappyPOD+R'' method uses the QR and random selection methods for $p\le r$ and $p>r$, respectively.
Comparison between the QR, GappyPOD+R, and random selection methods will help us to evaluate the performance of the QR method in the oversampling case. This ``GappyPOD+R'' method is exactly the same as the ``GappyPOD+R'' method in\cite{peherstorfer2020stability,clark2020multi}.

Peherstorfer \textit{et al.} proposed a sensor selection method which is based on the lower bounds of the smallest eigenvalues of certain structured matrix updates in the oversampling case, which is referred to as the ``GappyPOD+E'' method\cite{peherstorfer2020stability,clark2020multi}. It should be noted that this second method performed the best among the oversampling-point selection methods that they compared\cite{peherstorfer2020stability}.

\begin{figure}[htbp]
    \centering
    \includegraphics[width=3.3in]{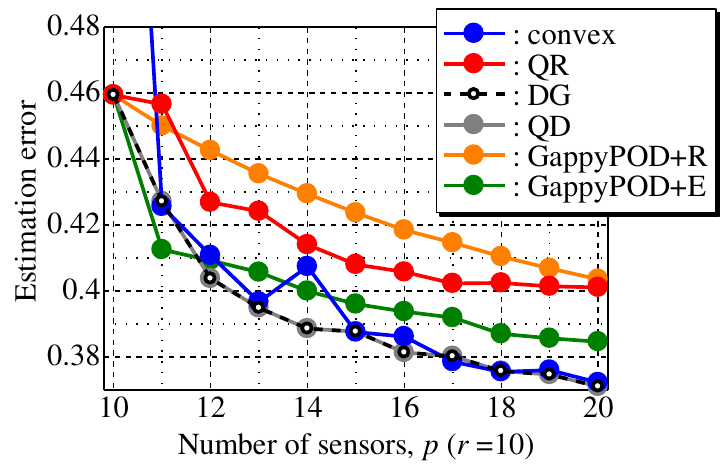}
    \caption {Estimation error against the number of sensors for the number of POD modes $r=10$ in the NOAA-SST problem.}
    \label{fig:estimation error oversampling}
\end{figure}

Fig. \ref{fig:estimation error oversampling} shows the relationship between the estimation error and the number of sensors for the number of POD modes $r=10$ in the NOAA-SST problem for the oversampling case ($p>r$) when the training and validation data are the same as each other. The estimation error of the GappyPOD+R method (solid orange line with closed circles) lies between those of the QR and random selection methods (see Fig. \ref{fig:EstimationError}). The QR method provides better sensor selection than the random selection method although the former has a huge computational cost in the oversampling case, as shown in Fig. \ref{fig:CalculationTime}. The estimation error of the GappyPOD+E method (solid green line with closed circles) is less than that of the QR method and is the smallest of all the methods in the case of $p=11$. However, the method proposed in this study, the DG method, performs better sensor selection in the oversampling case except for $p=11$ than does the GappayPOD+E method, which is almost the best method ever proposed among oversampling point selection methods in the DEIM framework.

\bibliographystyle{IEEEtran}
\bibliography{xaerolab}

\begin{IEEEbiography}[{\includegraphics[width=1in,height=1.25in,clip,keepaspectratio]{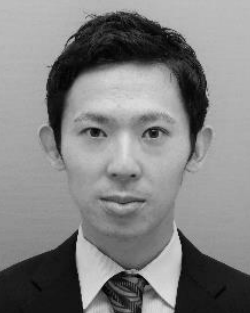}}]{Yuji Saito} received the B.S. degree in mechanical engineering, and the Ph.D. degree in mechanical space engineering from Hokkaido University, Sapporo, Japan in 2014 and 2018. He is an Assistant Professor of Aerospace Engineering at Tohoku University, Sendai, Japan.
\end{IEEEbiography}

\begin{IEEEbiography}[{\includegraphics[width=1in,height=1.25in,clip,keepaspectratio]{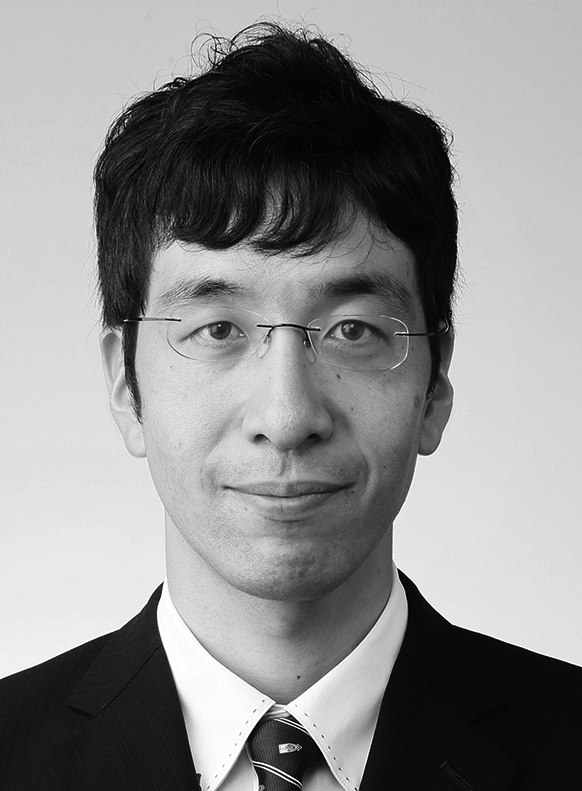}}]{Taku Nonomura} received the B.S. degree in mechanical and aerospace engineering from Nagoya University, Nagoya, Japan in 2003, and the Ph. D. degree in aerospace engineering from the University of Tokyo, Tokyo, Japan in 2008. He is currently an Associate Professor of Aerospace Engineering at Tohoku University, Sendai, Japan. 
\end{IEEEbiography}

\begin{IEEEbiography}[{\includegraphics[width=1in,height=1.25in,clip,keepaspectratio]{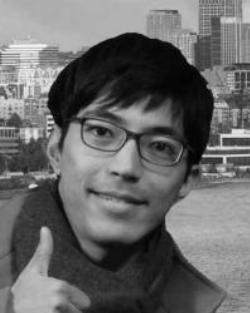}}]{Keigo Yamada} received the B.S. degree in physics from Tohoku University, Sendai, Japan, in 2019. He is a M.S. student of engineering at Tohoku University.
\end{IEEEbiography}

\begin{IEEEbiography}[{\includegraphics[width=1in,height=1.25in,clip,keepaspectratio]{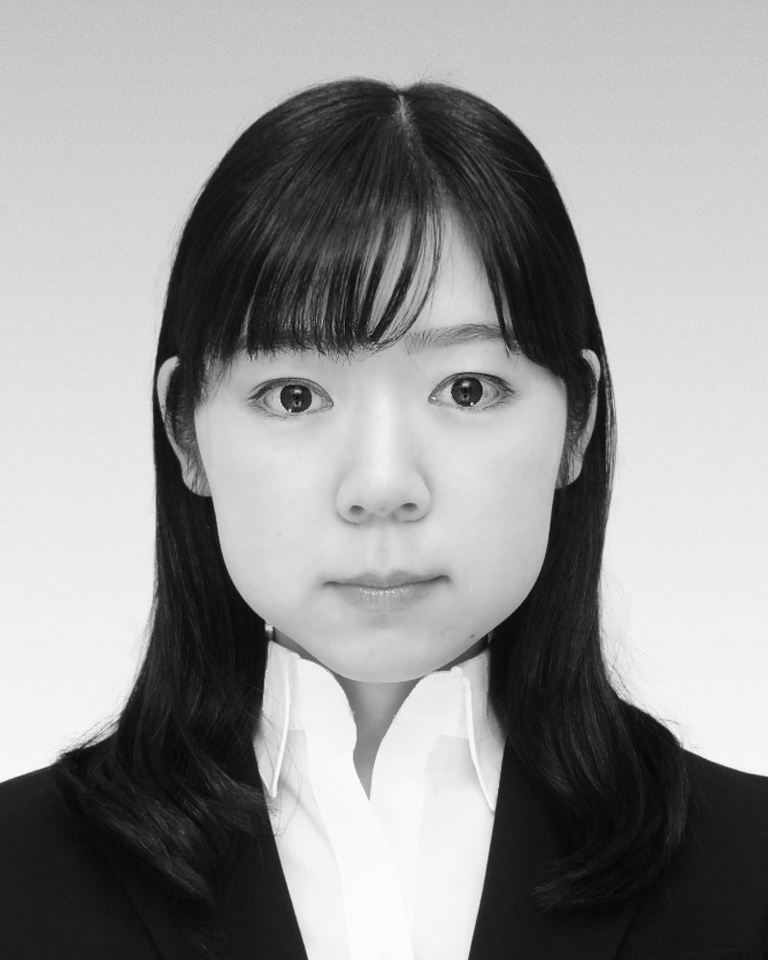}}]{Kumi Nakai} received the Ph.D. degree in mechanical systems engineering from Tokyo University of Agriculture and Technology, Japan, in 2020. From 2017 to 2020, she was a Research Fellow of the Japan Society for the Promotion of Science (JSPS) at Tokyo University of Agriculture and Technology, Japan. Since 2020, she has been a postdoctoral researcher at Tohoku University, Japan. Her research interests include data-driven science, discharge plasma dynamics, fluid dynamics, and flow control utilizing atmospheric pressure plasma.
\end{IEEEbiography}

\begin{IEEEbiography}[{\includegraphics[width=1in,height=1.25in,clip,keepaspectratio]{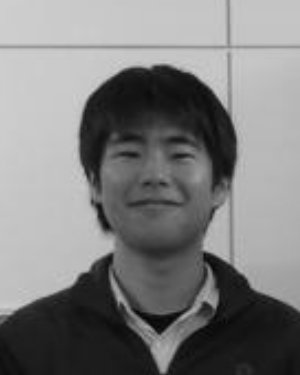}}]{Takayuki Nagata} received the B.S. and M.S. degrees in mechanical and aerospace engineering from Tokai University, Japan, in 2015 and 2017, respectively. He received the Ph.D. degree in aerospace engineering from Tohoku University, Japan, in 2020.  From 2018 to 2020, he was a Research Fellow of Japan Society for the Promotion of Science (JSPS) at Tohoku University, Japan. He is currently a postdoctoral researcher at Tohoku University, Japan. 
\end{IEEEbiography}

\begin{IEEEbiography}[{\includegraphics[width=1in,height=1.25in,clip,keepaspectratio]{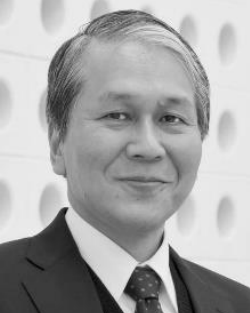}}]{Keisuke Asai} received the bachelor degree in aeronautical engineering from Kyoto University, Japan, in 1980, and the Ph.D. degree from the University of Tokyo in 1995. He worked at the National Aerospace Laboratory (now, Aeronautical Directorate of JAXA) from 1980 to 2003. He is currently a Professor of Aerospace Engineering at Tohoku University, Sendai, Japan.
\end{IEEEbiography}

\begin{IEEEbiography}[{\includegraphics[width=1in,height=1.25in,clip,keepaspectratio]{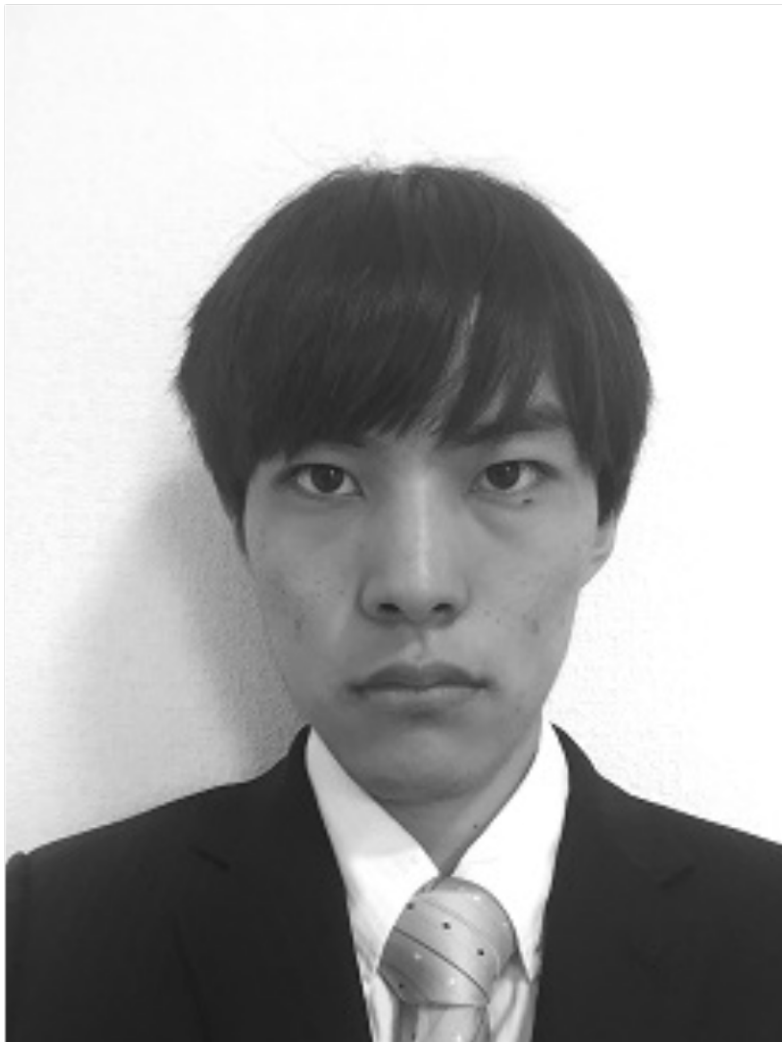}}]{Yasuo Sasaki} received the B.S. degree from Nagoya University, Nagoya, Japan in 2017. He is currently a Ph.D. student of engineering at Nagoya University.
\end{IEEEbiography}

\begin{IEEEbiography}[{\includegraphics[width=1in,height=1.25in,clip,keepaspectratio]{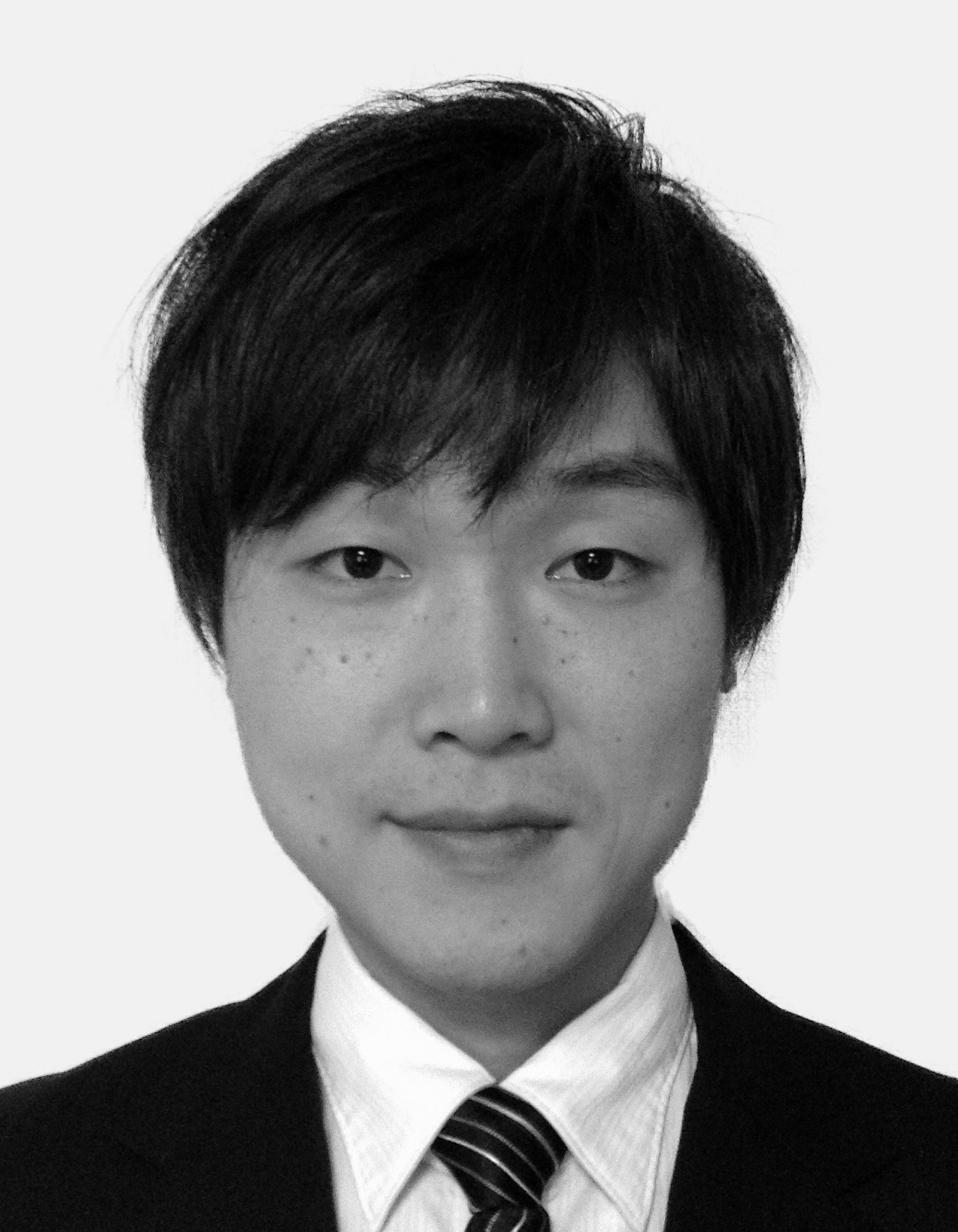}}]{Daisuke Tsubakino} received the B.S. degree from Nagoya University, Nagoya, Japan in 2005 and the Ph.D. degree in information science and technology from the University of Tokyo, Tokyo, Japan 2011. He is currently a lecturer at Nagoya University.
\end{IEEEbiography}

\EOD
\end{document}